\documentstyle[pslatex,epsf,epsfig,psfig]{mn2e}

\newif\ifAMStwofonts
\AMStwofontstrue

\newcommand{\beqn}{\begin{eqnarray*}}
\newcommand{\eeqn}{\end{eqnarray*}}

\newcommand{\p}{\partial}
\newcommand{\f}[2]{\frac{#1}{#2}}

\def\be{\begin{equation}}
\def\ee{\end{equation}}
\def\beq{\begin{eqnarray}}
\def\eeq{\end{eqnarray}}
\def\nn{\nonumber}

\begin{document}

\title[
Approximate 
Matching of Analytic and Numerical Solutions 
for Rapidly Rotating Neutron Stars]
{ Approximate Matching of Analytic and Numerical Solutions for 
Rapidly Rotating Neutron Stars}
\author[Emanuele Berti and Nikolaos Stergioulas]
{Emanuele Berti$^{1,2,3}$ and Nikolaos Stergioulas$^{1}$ \\
$^{1}$ Department of Physics, Aristotle University of Thessaloniki,
 Thessaloniki 54124, Greece\\
$^{2}$ McDonnell Center for the Space Sciences, Department of Physics,
Washington University, St. Louis, Missouri 63130, USA\\
$^{3}$ Present address: Groupe de Cosmologie et Gravitation
(GReCO), Institut d'Astrophysique de Paris (CNRS), $98^{bis}$ Boulevard
Arago, 75014 Paris, France
}

\maketitle

\begin{abstract}
  
  We investigate the properties of a closed-form analytic solution
  recently found by Manko {\it et al.} (2000b) for the exterior
  spacetime of rapidly rotating neutron stars.  For selected equations
  of state we numerically solve the full Einstein equations to
  determine the neutron star spacetime along constant rest mass
  sequences. The analytic solution is then matched to the numerical
  solutions by imposing the condition that the quadrupole moment of
  the numerical and analytic spacetimes be the same. For the analytic
  solution we consider, such a matching condition can be satisfied
  only for very rapidly rotating stars. When solutions to the matching
  condition exist, they belong to one of two branches.  For one branch
  the current octupole moment of the analytic solution is very close
  to the current octupole moment of the numerical spacetime; the other
  branch is more similar to the Kerr solution.  We present an
  extensive comparison of the radii of innermost stable circular
  orbits (ISCOs) obtained with a) the analytic solution, b) the Kerr
  metric, c) an analytic series expansion derived by Shibata and
  Sasaki (1998) and d) a highly accurate numerical code. In most cases
  where a corotating ISCO exists, the analytic solution has an
  accuracy consistently better than the Shibata-Sasaki expansion. The
  numerical code is used for tabulating the mass-quadrupole and
  current-octupole moments for several sequences of constant rest
  mass.
 
\end{abstract}


\section{Introduction}

The analytic description of the vacuum spacetime surrounding a rapidly
rotating neutron star is still an open problem. The analytic structure
of the spacetime outside a slowly rotating star, and its relation to
the Kerr metric, has been well understood since the seminal works of
Hartle (1968) and Hartle \& Thorne (1969). On the other hand,
numerical solutions of the Einstein equations for stars rotating up to
the mass-shedding limit are now routinely obtained with a number of
different methods, such as the Komatsu, Eriguchi and Hachisu (1989)
method (see Stergioulas 2003, for an extensive comparison of the
different existing numerical methods).  These numerical solutions are
indeed useful for modelling astrophysical systems, for studying linear
perturbations of rapidly rotating relativistic stars and as initial
data for dynamical evolutions of spacetimes in numerical relativity
(see e.g. Stergioulas \& Friedman 1998, Stergioulas, Kluzniak \& Bulik
1999, Stergioulas \& Font 2001).

Despite the availability of numerical solutions, a consistent analytic
representation of the vacuum metric outside a rapidly rotating neutron
star is desirable for several reasons.  In the first place, having an
analytic form for the metric simplifies the computation of the {\it
  stationary} properties of the spacetime.  For example, if an
accurate analytic solution were available, geodesics in the neutron
star exterior could be studied analytically, and one could find
closed-form expressions for the radii and frequencies of the innermost
stable circular orbits (ISCOs).  In turn, this would simplify the
calculation of properties of accretion disks, of epicyclic
frequencies, of accretion luminosities, and so on.

Furthermore, having an analytic solution could prove useful to the
study of {\it dynamical} properties of the spacetime, such as
gravitational wave emission. One of the unsolved problems in
gravitational-wave theory is the study of the quasinormal modes of
rapidly rotating neutron stars. These can be computed either in the
frequency domain, as an eigenvalue problem, or in the time domain,
evolving numerically the (linearized or full) Einstein equations and
then computing the outgoing radiation. The major technical issue in
this problem is related to the difficulty of imposing outgoing-wave
boundary conditions at infinity, since a rapidly rotating neutron star
spacetime is expected to deviate significantly from Petrov type
D. Having in hand an accurate analytic metric for the exterior
spacetime one could envisage the possibility of computing the Weyl
scalars in closed form, looking for neutron star models which are, in
some suitably defined sense, ``close to Petrov type D'' (Baker
\& Campanelli 2000). If the spacetime is ``close enough to type D''
one could then apply approximation schemes to impose the outgoing-wave
boundary conditions.  The idea here is to improve the presently
available methods, which are generally based on the use of the Zerilli
functions (see e. g. Abrahams {\it et al.} 1992, Allen {\it et al.}
1998, Rupright {\it et al.} 1998) - i.e., on perturbations of {\it
spherically symmetric} vacuum spacetimes. Only recently, the Teukolsky
formalism for perturbations of Kerr black holes has been used for the
purpose of wave extraction in the final phase of binary black holes
mergers (Baker {\it et al.} 2002).

Until the development of a powerful integral equation method, devised
by Sibgatullin in 1984 (see Sibgatullin 1991 and Manko \& Sibgatullin
1993 for details), finding analytic solutions to the Einstein
equations for stationary axisymmetric spacetimes was largely a matter
of guesswork. One typically had to choose some ansatz to simplify the
mathematical problem of obtaining the solution; then one verified {\it
a posteriori} that the obtained solution had physically acceptable
properties. In Sibgatullin's method one knows the physical
characteristics of the solution to be constructed from the very
beginning, through the choice of the axis expressions of the Ernst
potentials.

A complete analytic representation of axisymmetric spacetimes can be
obtained in terms of a series expansion whose coefficients are the
physical multipole moments (Fodor, Hoenselaers \& Perjes 1989, Ryan
1995). In principle, this gives an approximation to a numerical
spacetime that can be made arbitrarily accurate: one would need to
include a sufficiently large number of multipole moments and match
them to some given numerical solution. However, such a procedure involves a
very large number of expansion coefficients, which makes it difficult
to use for practical purposes. Some  applications of this
idea have already appeared: for example, Shibata \& Sasaki (1998)
derived formulae for the location of the ISCO around
rapidly rotating neutron stars. 

Quite recently, Manko {\it et al.} (2000b) were able to find a new
asymptotically-flat solution to the Ernst equations for the
Einstein-Maxwell system. This solution is very interesting because it
is given {\it in closed form}. Furthermore, when two of its parameters
(i.e., the charge and magnetic moment) are set to zero, the solution
depends only on {\it three parameters}: mass, angular momentum and a
third parameter $b$, which is related to the spacetime's physical
quadrupole moment. With this simplification, the solution reduces to a
particular three-parameter specialization of the Kinnersley-Chitre
(1978) solution (a generalization of the Tomimatsu-Sato $\delta=2$
spacetime). Notice however that Kinnersley and Chitre only constructed
the relevant Ernst potential (they did not provide explicit
expressions for the corresponding metric functions). Furthermore, the
Kinnersley-Chitre solution is restricted to the subextreme case
($M^2>a^2$). On the other hand, in the solution by Manko {\it et al.},
when electric and magnetic fields are set to zero $M$ and $a$ are
allowed to assume arbitrary real values, because the parameter set in
their solution is analytically extended. Therefore the
Kinnersley-Chitre solution is obtained as a particular case of the
analytic solution in Manko {\it et al.} (2000b) when certain
restrictions are imposed on the parameters of that solution.

There have been attempts in the literature to fix the free parameters
in analytic exterior solutions by matching them to numerical
solutions.  However, different matching conditions were used.  For
example, Sibgatullin \& Sunyaev (1998, 2000) fixed the free parameters
appearing in a different analytic solution using the radii of
marginally stable circular orbits, or a suitably defined redshift
parameter at the stellar equator. For their metric, which is different
from the one we consider here, they found that corrections due to the
quadrupole moment can accurately reproduce the properties of the
``exact'' exterior spacetime only for several equations of state
(EOSs), with the exception of EOSs with large phase transitions. A
simple, closed form expression for the analytic metric used in
Sibgatullin \& Sunyaev (1998, 2000) was given explicitly by
Sibgatullin (2002).

A matching procedure based on the redshift parameter was again used by
Stute \& Camenzind (2002). Our own preference here is to avoid
matching using {\it local} properties and, instead, match the
solution's mass-quadrupole moment, which is a {\it global} property of
the spacetime. Furthermore, it is well known that deviations from the
slow-rotation behavior in rapidly rotating stars, due to the stellar
oblateness, are determined mainly by the mass-quadrupole moment. The
quadrupole moment was also used in matching the analytic and
numerical solution in Manko {\it et al.} (2000a). 

The plan of the paper is as follows. In section \ref{numgravfield} we
describe the procedure to numerically compute the spacetime describing
a rapidly rotating compact star using the Komatsu-Eriguchi-Hachisu
(1989) self-consistent field method, as modified by Cook, Shapiro and
Teukolsky (1994, henceforth CST). In particular, we discuss how to
implement this method for a numerical evaluation of the spacetime's
multipole moments.  In section \ref{analgravfield} we present the
analytic solution recently obtained by Manko {\it et al.} (which is
only valid in the vacuum prevailing outside the rotating neutron star)
and describe its multipolar structure.  In section \ref{match} we
describe our procedure to match Manko's analytic solution to the
numerically obtained spacetime, and derive the coordinate
transformation relating the two metrics.  Section \ref{checkSol} is
devoted to a discussion of the tests we used in order to understand
``how close'' the analytic and numerical spacetimes are. As we will
discuss in the following, there are two possible families of analytic
solution for which the mass-quadrupole moment of the analytic solution
matches to the mass-quadrupole moment of the numerical spacetime. The
current-octupole moment of the first family of solutions is very close
to the current-octupole moment of the numerical spacetime, while the
second solution is close to the Kerr spacetime. An examination of the
metric functions on the equatorial plane and on the rotation axis
confirms that the first solution is also the one which better
approximates the numerically obtained metric functions.  As an
independent check, we compute the location of ISCOs in the spacetime
surrounding the rotating star using different approaches. In
particular we locate ISCOs using the analytic solution, and compare
the results thus obtained: 1) to the ISCOs found by numerical
integration of the Einstein equations, and 2) to the analytic formulae
for the ISCO's obtained by Shibata \& Sasaki (1998), truncated at
different orders of approximation. In most cases where a corotating
ISCO exists, the analytic solution has an accuracy consistently better
than the Shibata-Sasaki expansion. Only in some cases the higher-order
multipoles that are missing in the analytic solution significantly
increase the error in computing the location of the ISCO.  Finally, we
compare our matching procedure to previous work by Manko {\it et al.}
(2000a) and by Stute \& Camenzind. The conclusions follow.

\section{Numerical gravitational field of a rapidly rotating neutron star}
\label{numgravfield}

To begin with, in this section we briefly discuss the procedure for
obtaining highly-accurate numerical solutions for the spacetime of
rapidly rotating neutron stars and for computing their multipole
moments. For more details the reader is referred to the review article
by Stergioulas (2003).

\subsection{Numerical determination of the spacetime and computation of the multipole moments}
\label{numsol}

The interior and exterior spacetime of a stationary, axisymmetric star
is described by a metric in the following form:
\beq
\label{CSTmetr}
ds^2=-e^{2\nu}dt^2+B^2 e^{-2 \nu}r^2\sin^2\theta (d\phi-\omega dt)^2
+e^{2\alpha}(dr^2+r^2d\theta^2),
\eeq
where $\nu,~B, ~\alpha$ and $\omega$  are four metric functions to be determined by solving four
field equations. In the numerical method of Komatsu {\it et al.} (1989, 
henceforth KEH) one defines two auxiliary functions $\bar \rho$, $\bar \gamma$ through the
relations $\nu=(\bar \gamma + \bar \rho)/2$ and $B=e^{\bar \gamma}$.
Then, three out of the four field equations are written in the following
integral forms
\beq
\nn
&&\bar \rho(r,\mu)=-\sum_{n=0}^\infty e^{-\bar \gamma/2}
\\
&&\left\{
\int_0^\infty dr'\int_0^1 d\mu'
r'^2 f_{2n}^2(r,r') P_{2n}(\mu')S_{\bar \rho}(r',\mu')
\right\}\nn\\
&&P_{2n}(\mu),
\label{rho}
\eeq

\beq
\nn
&&\bar \gamma(r,\mu)=-{2\over \pi r\sin\theta}\sum_{n=1}^\infty e^{-\bar \gamma/2}
\\
&&\left\{
\int_0^\infty dr'\int_0^1 d\mu'
r'^2
{f_{2n-1}^1(r,r') \sin[(2n-1)\theta']\over 2n-1}
S_{\bar \gamma}(r',\mu')
\right\}\nn\\
&&\sin[(2n-1)\theta],
\eeq

\beq
\nn
&&\omega(r,\mu)=-{1\over r\sin\theta}\sum_{n=1}^\infty e^{(2\bar \rho-\bar \gamma)/2}
\\
&&\left\{
\int_0^\infty dr'\int_0^1 d\mu'
r'^3 {\sin\theta'f_{2n-1}^2(r,r')\over 2n(2n-1)}
P_{2n-1}^1(\mu')S_\omega(r',\mu')
\right\}\nn\\
&&P_{2n-1}^1(\mu) ,
\label{omega}
\eeq
where
\beq
f_n^1(r,r')&=&\left({r'\over r}\right)^n\qquad\hbox{for}\qquad r'\leq r,
\nn\\
f_n^1(r,r')&=&\left({r\over r'}\right)^n\qquad\hbox{for}\qquad r'>r,
\nn\\
f_n^2(r,r')&=&{1\over r}\left({r'\over r}\right)^n\qquad\hbox{for}\qquad r'\leq r,
\nn\\
f_n^2(r,r')&=&{1\over r}\left({r\over r'}\right)^n\qquad\hbox{for}\qquad r'>r,
\nn
\eeq
and $S_{\bar \rho}$, $S_{\bar \gamma}$ and $S_\omega$ are lengthy
source terms, whose expressions can be found in KEH. In the equations
above, $\mu=\cos\theta$, while $P_n(\mu)$ denotes the Legendre
polynomials and $P_n^m(\mu)$ the associated Legendre functions. The
metric function $\alpha$ is determined by an ordinary differential
equation.

We compute numerical equilibrium models using the code by Stergioulas
and Friedman (1995) (see Nozawa {\it et al.} 1998 and Stergioulas 2003 for
extensive accuracy tests). The numerical code uses the CST
formulation, in which the KEH equations are written in terms of a
compactified coordinate $s$ defined through the relation
\be
r=r_e\left(s\over 1-s\right), 
\ee
where $r_e$ is the (coordinate) radius of the stellar equator.  This
allows the computation of the whole exterior spacetime out to
infinity, which is important in detailed comparisons of the numerical
metric to the analytic metric.

For a configuration that is stationary, axisymmetric, symmetric with
respect to reflections in the equatorial plane and asymptotically
flat, the spacetime can be characterized by two sets of scalar
multipole moments: the even-valued mass moments ($M_0,~M_2,~M_4\dots$)
and the odd-valued current moments ($S_1,~S_3,~S_5\dots$).  Ryan
(1997) presented a method for extracting the multipole moments from
the asymptotic form of the metric functions. The lowest-order
appearance of each moment in terms of a power series in $1/r$ is
determined by the expansions \be \bar
\rho=\sum_{n=0}^\infty~-2{M_{2n}\over r^{2n+1}}P_{2n}(\mu), \label{rhotilde}
\ee and
\be \omega=\sum_{n=1}^\infty{-2\over 2n-1}{S_{2n-1}\over r^{2n+1}}
{P_{2n-1}^1(\mu)\over \sin\theta}. \label{omegatilde} \ee 
By comparison with equations
(\ref{rho}) and (\ref{omega}) one finds that \be\label{Mn}
M_{2n}={1\over 2}\int_0^\infty dr'\int_0^1 d\mu' r'^{2n+2}
P_{2n}(\mu')S_{\bar \rho}(r',\mu'), \ee and \be\label{Sn}
S_{2n-1}={1\over 4n}\int_0^\infty dr'\int_0^1 d\mu' r'^{2n+2}
\sin\theta' P_{2n-1}^1(\mu')S_\omega(r',\mu').  \ee Thus, in general,
any model of a rapidly rotating neutron star has an infinite number of
mass- and current-multipole moments. In order to match an analytic
exterior metric to a numerically-computed interior metric and to check
the accuracy of the matching procedure, we  computed the
mass-quadrupole moment $M_2$ and the current-octupole moment $S_3$.

An alternative, asymptotic method for evaluating the multipole moments
was introduced by Laarakkers \& Poisson (1997). We also used their method
in order to cross-check the results obtained from the integral relations
(\ref{Mn}) and (\ref{Sn}).  The idea, in this case, is to evaluate
numerically the coefficient of $P_{2n}(\mu)$ in the general expression
for (\ref{rho}) - or analogously, the coefficient of $P_{2n-1}^1$ in
(\ref{omega}) - at the outermost grid points (i.e., as $r\to \infty$),
and multiply the result by the appropriate factor (containing powers
of $r$) that can be obtained from equations (\ref{rhotilde}) and
(\ref{omegatilde}).  We have checked that the two methods typically
agree to better than one part in $10^3$.

 \begin{figure} 
   \psfig{figure=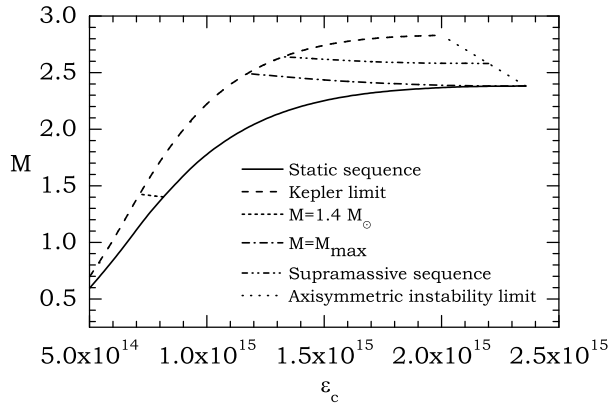,width=8cm}~~~~~ \caption{ Limit sequences
     for EOS APR: the solid line corresponds to the
     nonrotating limit; the long-dashed line corresponds to the
     mass-shedding (Kepler) limit; the dotted line is the axisymmetric
     instability limit. The nearly horizontal lines are sequences of
     constant rest mass. From bottom to top: the dashed line
     corresponds to a star of gravitational mass $M=1.4 M_\odot$ in
     the non rotating limit; the dash-dot line is a maximum-mass
     normal sequence; and finally, the dash-dot-dot line is a selected
     supramassive sequence.}
 \label{StaticKepler}
 \end{figure}

\subsection{Equilibrium sequences}

The equilibrium solutions for a given EOS form a two-parameter
family. In particular, {\it stable} equilibrium solutions are bounded
by four limit sequences. These limits are shown in Figure
\ref{StaticKepler}, which displays the gravitational mass $M$ vs.  the
central energy density $\epsilon_c$ for one the EOSs derived by Akmal,
Pandharipande and Ravenhall (1998, henceforth APR). As an illustrative
example we consider the APR EOS which does not include boost
interactions (we refer to the original paper for details). The
qualitative picture does not change when we consider other EOSs (see
eg. CST, where plots are presented for a representative sample of
EOSs). The solid line is the {\it static} limit - that is, the
sequence of nonrotating solutions to the standard
Tolman-Oppenheimer-Volkoff equations. The long-dashed line is the {\it
mass-shedding} (Kepler) limit, which is determined by the condition
that the centrifugal force exactly balances the gravitational
attraction at the stellar equator, in which case a fluid element on
the equator has the same angular velocity as a free particle in a
Keplerian orbit at the same location. Both sequences terminate at high
central density at the {\it stability} limit, where equilibrium
solutions are marginally stable to axisymmetric perturbations; and
they terminate at low central densities at the {\it low-mass} limit,
below which a neutron star cannot form (not shown in Figure
\ref{StaticKepler}).

Within the class of stable equilibrium solutions, CST pointed out the
significance of constant rest-mass sequences, called {\it evolutionary
  sequences}, since an isolated neutron star, slowly losing energy and
angular momentum via some dissipative process (e.g.  electromagnetic
emission or gravitational-wave radiation), must evolve conserving the
total baryon number, and hence its rest mass $M_B$. An accreting
neutron star in a binary system will not evolve along a constant
rest-mass sequence: the actual sequence depends on several parameters,
such as the magnetic field, accretion rate etc.  Nevertheless, the
constant rest-mass sequences in CST have been used in the past in
evaluating the accuracy of analytic exterior solutions and we will
also use them here solely for the same reason. We compute three
constant rest-mass sequences for each EOS:
\begin{itemize}
\item
the sequence corresponding to a canonical neutron star having
gravitational mass $M=1.4 M_\odot$ in the non rotating limit,

\item
the sequence terminating at the maximum-mass model in the non 
rotating limit (maximum-mass normal sequence).

\item
a supramassive sequence, i.e., a sequence which does not terminate at
a nonrotating model.
\end{itemize}

We include the following set of EOSs. For comparison with CST we
include EOSs A, AU, FPS and L. We refer to their paper for an extensive
discussion of each EOS.  We supplement the set of EOSs considered by
CST with a relatively new model: the model derived by Akmal,
Pandharipande and Ravenhall (1998) from Hamiltonian many-body theories
of nuclear matter, including boost corrections in the Hamiltonian
(henceforth, we will refer to this model as APR-b, where ``b'' stands
for ``boosted'').

In Tables (\ref{EOSA}--\ref{EOSAPRb}) we give numerical results for
the structure properties of the models we have computed.  All models
have been computed using a resolution of ($301$ angular points)
$\times$ ($601$ radial points), corresponding to a typical accuracy of
at least one part in $10^3$ in all quantities.  Each Table corresponds
to a constant rest mass sequence, and lists: the total central energy
density $\epsilon_c$ in units of $10^{15}$ g~cm$^{-3}$; the angular
velocity $\Omega$ in units of $10^3$ s$^{-1}$; the moment of inertia
$I$ in units of $10^{45}$ g~cm$^2$ (for rotating models only); the
gravitational mass $M$ in solar masses; the ratio of rotational
kinetic energy to gravitational binding energy $T/W$; the equatorial
circumferential radius of the star $R_e$ in km and the height (in km)
of corotating ($h_+$) and counterrotating ($h_-$) ISCOs from the
surface of the star (if an ISCO does not exist, the corresponding
entry is omitted). The height of an ISCO is defined as the difference
between the circumferential radius at the ISCO and the circumferential
equatorial radius of the star.  The next three columns give the first
few physical multipoles in geometrized units of $c=G=1$: namely, we
list the mass quadrupole moment $M_2\equiv Q$ in km$^3$, the angular
momentum $S_1 \equiv J$ in km$^2$, and the current octupole moment $S_3$ in
km$^4$ .  We have checked our code by reproducing the quadrupole
moments computed by Laarakkers \& Poisson and found excellent
agreement. The accuracy in computing $S_3$ was checked by
comparing the integral form to the asymptotic form mentioned in Section
\ref{numsol}, finding good agreement.  This shows that
using the compactified coordinate introduced in CST allows a very
accurate numerical determination of relatively high-order multipoles.

The last column gives the value of the ``quadrupole'' parameter $b$
(in km) for which the analytic solution provides a good approximation
of the numerical spacetime. When matching the quadrupole moment of the
numerical and analytic spacetimes is not possible, the corresponding
entry is omitted. More details on the procedure we followed to obtain
the values listed in this column will be given in section \ref{match}.

\section{Analytic gravitational field of a rapidly rotating neutron star}
\label{analgravfield}

In this section we summarize the properties of the vacuum analytic
solution obtained by Manko {\it et al.} (2000b). We will concentrate in
particular on the multipolar structure of the solution, since our
ultimate purpose will be to reproduce accurately the first few
multipoles of the numerical spacetimes we discussed in the previous
section.

\subsection{The solution by Manko {\it et al.}}                  
\label{mankosol}

In the vacuum region surrounding a stationary and axisymmetric star,
the spacetime only depends on three metric functions (while four
metric functions are needed for the interior). The most general form
of the metric (Papapetrou 1953) is
\beq
\label{Manko}
ds^2=-f(dt-wd\phi)^2+f^{-1}\left\{
e^{2\gamma} (d\tilde\rho^2+d\tilde z^2)+\tilde\rho^2d\phi^2
\right\}.
\eeq
Here $f$, $w$ and $\gamma$ are functions of the
quasi-cylindrical Weyl-Lewis-Papapetrou coordinates $(\tilde
\rho,~\tilde z)$.  Starting from this metric one can write down the
vacuum Einstein-Maxwell equations as two equations for two complex
potentials ${\cal E}$ and $\Phi$, following a procedure due to Ernst
(1968).  The equations are:
\beq
(Re \{{\cal E}\}+|\Phi|^2)\nabla^2{\cal E}&=&
(\nabla {\cal E}+2\Phi^*\nabla\Phi)\cdot\nabla{\cal E}\\
(Re \{{\cal E}\}+|\Phi|^2)\nabla^2\Phi&=&
(\nabla {\cal E}+2\Phi^*\nabla\Phi)\cdot\nabla\Phi
\eeq
Once the potentials are known, the metric can be reconstructed.
Sibgatullin (1991) devised a powerful procedure for reducing the
solution of the Ernst equations to simple integral
equations. Basically, one starts with a choice for the values of the
Ernst potentials on the symmetry axis
\be
e(\tilde z)\equiv{\cal E}(\tilde\rho=0,\tilde z), \qquad
f(\tilde z)\equiv\Phi(\tilde\rho=0,\tilde z),
\ee
solves two complex-valued integral equations, and checks that the obtained
solution satisfies the expression of the Ernst potentials in terms of physical
multipoles:
\beq
&&{\cal E}={1-\xi \over 1+\xi}, \qquad \Phi={q\over 1+\xi}\\
&&\xi(\tilde\rho=0)=\sum_{n=0}^\infty m_n \tilde z^{-(n+1)},\\
&&q(\tilde\rho=0)=\sum_{n=0}^\infty q_n \tilde z^{-(n+1)}
\eeq
The real parts of $m_n$ are the mass multipoles, the imaginary parts of $m_n$
are the current multipoles, the real parts of $q_n$ are the electric multipoles
and the imaginary parts of $q_n$ are the magnetic multipoles.

After more than ten years of work in the field, Manko {\it et al.} (2000b)
were finally able to find a vacuum solution involving five parameters
(mass, angular momentum, charge, magnetic dipole moment and mass
quadrupole moment) which can be expressed in terms of relatively
simple rational functions.  We are particularly interested in
solutions having no charge or magnetic dipole moment.  If we denote
by $M$ the gravitational mass of the star, by $a$ the specific angular
momentum ($a=J/M$), and introduce a parameter $b$ which can be related
to the mass quadrupole moment, their choice for the axis values of the
Ernst potentials is: 
\begin{equation} 
 e(z)={(z-M-ia)(z+ib)+d-\delta-ab \over
  (z+M-ia)(z+ib)+d-\delta-ab}, \qquad f(z)=0 
\end{equation}
with 
\beq
\delta&=&{-M^2b^2\over M^2-(a-b)^2},\\
d&=&{1\over 4}\left[M^2-(a-b)^2\right].  
\eeq 
To be able to write the metric in rational form, one must introduce generalized
spheroidal coordinates
\begin{equation}
x=\frac{r_++r_-}{2k}, \qquad
y=\frac{r_+-r_-}{2k},
\end{equation}
where $r_\pm=\sqrt{\tilde\rho^2+(\tilde z\pm k)^2}$ and 
\be 
k=\sqrt{d+\delta}.  
\ee 
The inverse transformation between the two sets of coordinates is 
\be 
\tilde\rho=k(1-y^2)^{1/2}(x^2-1)^{1/2},\qquad
\tilde z=kxy. \ee 
The metric is then written as 
\begin{eqnarray}
\label{solution}
ds^2 &=& f(dt-wd\phi)^2-\frac{k^2}{f}
\Bigl[
e^{2 \gamma}
(x^2-y^2)\left({dx^2\over x^2-1}+{dy^2\over 1-y^2}\right) \nn\\
&&+
(x^2-1)(1-y^2)d\phi^2
\Bigr],
\end{eqnarray}
with
\be
f={E\over D}, \qquad
e^{2 \gamma}={E\over 16k^8(x^2-y^2)^4}, \qquad
w={-(1-y^2)F\over E},
\ee
and
\beq
D&=&
\{4 (k^2x^2-\delta y^2)^2+ 2kmx[ 2k^2(x^2-1)\nn\\
&+&(2\delta+ab-b^2)(1-y^2) ]\nn\\
&+&(a-b)[ (a-b)(d-\delta)-m^2b ](y^4-1)-4d^2 \}^2 \nn\\
&+&4y^2\{2k^2(x^2-1)[kx(a-b)-mb]-2mb\delta(1-y^2)\nn\\
&+&[ (a-b)(d-\delta)-m^2b ](2kx+m)(1-y^2) \}^2 \\
&& \nn\\
E&=&
\{4 [k^2(x^2-1)+\delta(1-y^2)]^2\nn\\
&+&(a-b)[ (a-b)(d-\delta)-m^2b ](1-y^2)^2\}^2\nn\\
&-& 16k^2(x^2-1)(1-y^2)\{ (a-b)[k^2(x^2-y^2)+2\delta y^2]\nn\\
&+&m^2by^2 \}^2 \\
&&\nn\\
F&=&
8k^2(x^2-1)\{ (a-b)[k^2(x^2-y^2)+2\delta y^2 ]+y^2m^2b \}\nn\\
&\times&
\{ kmx[ (2kx+m)^2-2y^2(2\delta+ab-b^2)-a^2+b^2 ]\nn\\
&-&2y^2(4\delta d-m^2b^2) \}
+\{ 4[k^2(x^2-1)+\delta(1-y^2)]^2 \nn\\
&+& (a-b)[ (a-b)(d-\delta)-m^2b ](1-y^2)^2\} \nn \\
&\times&
(4(2kmbx+2m^2b)[k^2(x^2-1)+\delta(1-y^2)]\nn\\
&+&(1-y^2)\{ (a-b)(m^2b^2-4\delta d)\nn\\
&-&(4kmx+2m^2)[ (a-b)(d-\delta)-m^2b ] \} ).
\eeq
In order for the solution to satisfy the requirements of axisymmetry,
stationarity and reflection-symmetry in the equatorial plane, all
three parameters $M$, $a$ and $b$ must be real.

\subsection{Multipolar structure of the analytic solution}

Here we examine the multipolar structure of the analytic solution by
Manko {\it et al.}  for rotating and nonrotating solutions. The only
nonvanishing multipole moments of the solution are the gravitational
mass $Re\{m_0\}\equiv M$, the quadrupole moment $Re\{m_2\} \equiv Q$,
the angular momentum $Im\{m_1\} \equiv J=aM$ and the
current octupole $Im\{m_3\}=S_3$.  The quadrupole moment and the
current octupole moment are given in terms of the three parameters $M$, 
$a$ and $b$ as
\be
\label{quadanal} Q=-M(d-\delta-ab+a^2),  
\ee 
and
\begin{equation}
\label{S3}
S_3=-M \left \{ a^3 -2a^2b+a\left[b^2+2(d-\delta)\right]-b(d-\delta) \right\}.
\end{equation}
However, since $a$ and $b$ are independent parameters, setting $a$
equal to zero does not automatically imply a vanishing $Q$ and $S_3$,
as would be the case for a realistic solution of a nonrotating perfect
fluid star. Instead, the nonrotating solution ($a=0$) has a quadrupole
moment equal to
\begin{equation}
Q(a=0)=-{M \over 4} {\left( M^2+b^2 \right)^2 \over \left( M^2-b^2 \right)},
\end{equation}
and a current octupole moment equal to 
\begin{equation}
S_3(a=0)=-bQ(a=0).
\end{equation}
It is obvious that there exists no real value of the parameter $b$ for
which the quadrupole moment vanishes for a nonrotating star. For $|b|<M$,
the solution is oblate ($Q<0$) with a minimum quadrupole deformation obtained
for $b=0$
\begin{equation}
|Q|_{\rm min}(a=0)=M^3/4.
\end{equation}
At $b=\pm M$, the nonrotating multipole moments $Q$ and $S_3$ diverge, while for $|b|>M$, 
the nonrotating solution is prolate ($Q>0$) with a minimum quadrupole deformation
of 
\begin{equation}
Q_{\rm min}(a=0)=2M^3,
\end{equation}
at $b=\pm \sqrt{3}M$.

Obviously, the analytic solution by Manko {\it et al.} does not reduce
continuously to the Schwarzschild solution as the rotation
vanishes. It can only reduce to other forms of nonrotating vacuum
solutions (e.g. the well-known Weyl solutions) that could be
matched to other interior solutions, such as nonrotating stars with
non-isotropic stresses, inducing nonvanishing quadrupole deformations.
Nevertheless, as we will show next, the analytic solution can
approximately describe a rapidly rotating fluid star, when the
rotation rate is large enough, so that the quadrupole deformation induced
by the rotation roughly exceeds the minimum nonvanishing oblate
quadrupole deformation of the solution in the absence of rotation,
i.e. roughly when
\begin{equation}
|Q| > M^3/4.
\end{equation}
Since the quadrupole moment is roughly
proportional to $a^2M$, one expects that the analytic solution could
be relevant for rotation rates of roughly $j>0.5$, where $j\equiv
J/M^2$ is a dimensionless measure of the angular momentum of the
star. We will confirm this expectation by direct comparisons with
numerical solutions in the next section.  

It is interesting that the Kerr  solution can still be obtained from 
the analytic solution, if one accepts the following imaginary form for the 
parameter $b$
\be\label{Kerrlimit2}
b=i\sqrt{M^2-a^2},
\ee
with $a \leq M$. In this case, one recovers the correct 
expressions $Q=-a^2M$ and $S_3=-a^3M$.

\begin{figure}
  \psfig{figure=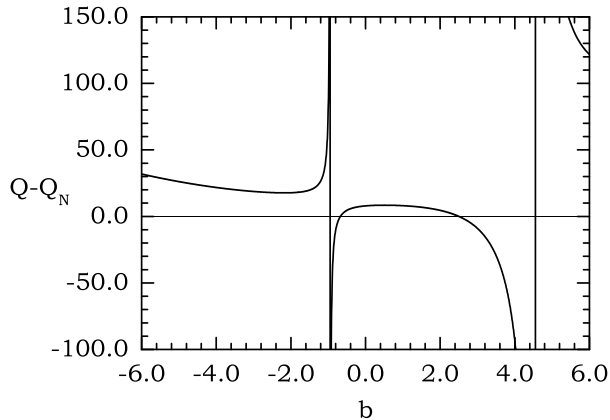,width=8cm}~~~~~ \caption{ The matching
    condition $Q-Q_{\rm N}=0$ as a function of $b$ for the fastest
    rotating FPS model in the maximum-mass sequence (last row of
    $M_B=2.105 M_\odot$ sequence in Table \ref{EOSFPS}). There are two
    possible solutions: a solution for which (typically, as in the
    case shown) $b=b_-<0$, and a positive solution for which
    $b=b_+>0$. The ``negative'' solution $b=b_-$ is the one which is
    relevant for rapidly rotating neutron stars, as shown in section
    \ref{checkSol}.  }
\label{analytic_bf}
\end{figure}

\begin{figure}
  \psfig{figure=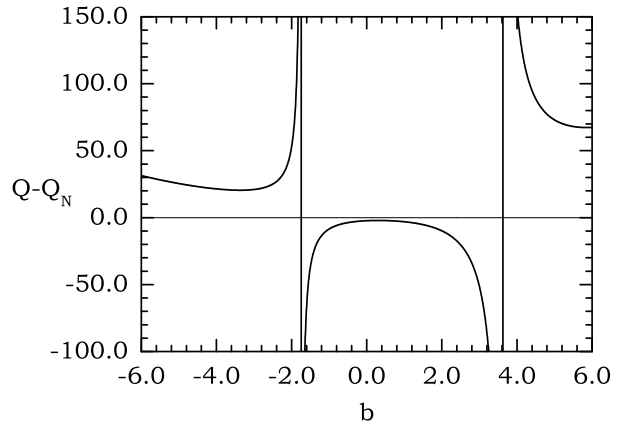,width=8cm}~~~~~ \caption{ The matching
    condition $Q-Q_{\rm N}=0$ as a function of $b$ for a slowly
    rotating model in the maximum-mass sequence for EOS FPS (fourth
    row of $M_B=2.105 M_\odot$ sequence in Table \ref{EOSFPS}). No real
    solution exists. }
\label{analytic_bs}
\end{figure}

\section{Matching the interior and exterior solutions}
\label{match}

The three parameters of the analytic solution ($M, a$ and $b$) can be
set at will. However, only certain combinations of values can
correspond to specific models of rapidly rotating neutron stars. Since
the solution has four nonvanishing multipole moments, but only three
free parameters, one can at most match three multipole moments of any
given numerical solution. The fourth multipole moment will then be
determined by the analytic solution, and its relative difference with
the (known) numerical value will be  a measure of the accuracy of
the analytic solution.  The four multipole moments are not equally
important for specifying a solution, $S_3$ being the least important,
even for the most rapidly rotating models. Therefore we choose to
match the analytic exterior solution to known numerical solutions by
matching the gravitational mass $M$, the specific angular momentum $a$
and the mass-quadrupole moment $Q$. One then hopes that the resulting
analytic solution will yield a value for the current-octupole moment
$S_3$ that is close to the corresponding value in the numerical
model. As we will show, there exists a branch of solutions for which
this is indeed the case. Manko {\it et al.} (2000a) also used the quadrupole
moment to match numerical and analytic solutions, but their examples
correspond to the other branch of solutions, for which the analytic
value of $S_3$ does not agree well with the numerical value.

For a given model of a rapidly rotating neutron star, we first
construct a highly accurate numerical solution, as described in
section \ref{numgravfield}. In the analytic solution (\ref{solution}),
we set $M$ and $a$ to be equal to the obtained numerical values. The
remaining parameter $b$ is then determined by solving the equation
\begin{equation}
\label{Qmatch}
Q-Q_{\rm N}=0,
\end{equation}
where $Q_{\rm N}$ is the value of the quadrupole moment obtained by
the numerical code. A plot of $Q-Q_{\rm N}$ as a function of the
parameter $b$, for the most rapidly rotating model of the maximum-mass
sequence for EOS FPS, is shown in Fig. \ref{analytic_bf}. Two possible
real solutions for $b$ exist: a solution that is usually negative,
$b_-$, and a solution that is always positive, $b_+$. Thus, for each
set of physical parameters $M, a$ and $Q$, there exists two different
branches of solutions, with parameters ($M$, $a$, $b_-$) and ($M$,
$a$, $b_+$), respectively.  In the remainder of the paper, we will
refer to these two different branches as the negative solution (-) and
the positive solution (+). As we will show next, these two branches
correspond to very different spacetimes.

In the previous section, we estimated that the analytic solution
should be relevant for rapidly rotating neutron stars only for values
of $j$ roughly larger than 0.5. Fig. \ref{analytic_bs} shows a more
slowly rotating model than the model shown in Fig. \ref{analytic_bf},
along the same evolutionary sequence. It is obvious that no solution
to equation (\ref{Qmatch}) exists for any real value of the parameter
$b$. Along each sequence there is a critical rotation rate above which
one can match the numerical interior solution to the analytic exterior
solution. In Tables (\ref{EOSA}--\ref{EOSAPRb}) we list all computed
physical properties for the selected sequences. The last column lists
the parameter $b=b_-$ of the {\it negative} branch of the analytic
solution (when it exists). This is the relevant branch for rapidly
rotating neutron stars, as we will show in section \ref{checkSol}.
In Tables (\ref{EOSA}--\ref{EOSAPRb}) some models appear having
$b=b_->0$. These models do {\it not} belong to the positive branch
$b_+$. They are instead models which are very close to the critical
value of the rotation parameter, $j=j_{crit}$: in these particular
cases, both solutions to equation (\ref{Qmatch}) can happen to be
positive.  However, in general (as long as a model is rotating
somewhat above the critical rate) the negative branch has $b_-<0$.

Typical values of $j_{crit}$ above which the analytic solution exists
are listed in Table \ref{jcrit} for a subset of the considered EOSs.
For smaller masses, $j_{crit}$ is usually smaller: therefore, for
``canonical'' neutron stars (having mass $M\sim 1.4 M_\odot$ in the
non rotating limit) the analytic solution is valid over a wider range
of $j$.  In terms of the angular velocity at the mass-shedding limit
for uniformly rotating stars, the critical rotation rates are given in
the right column of Table \ref{jcrit}. The critical rotation rate
$\Omega_{crit}/\Omega_{Kepler}$ ranges from $\sim 0.4$ to $\sim 0.7$
for the $M=1.4M_\odot$ sequence, with the lower ratio corresponding to
the stiffest EOS. For the maximum-mass sequence the ratio is $\sim
0.9$, nearly independent of the EOS. In conclusion, the analytic
exterior solution can be useful for studying rapidly rotating neutron
stars.  The exterior gravitational field of massive neutron stars
created in binary neutron star mergers, supported temporarily by
differential rotation against collapse, could also be described, to
some accuracy, by the analytic solution (the accuracy will depend on
how significant the higher multipole moments are in the case of strong
differential rotation). If the EOS is very stiff, such as EOS L, then
the analytic solution is also valid for for description of accreting
neutron stars in Low-Mass-X-Ray binaries (LMXB), with rotational
periods of a few milliseconds.

\subsection{Coordinate transformations between vacuum and non-vacuum metrics}

Before presenting specific tests of the accuracy of the analytic solution, 
we need to describe the coordinate transformation that relates
the interior metric  (\ref{CSTmetr}) to the exterior {\it vacuum} metric 
(\ref{Manko}) (see Islam 1985) .  For  the interior metric (\ref{CSTmetr}), 
we define the cylindrical coordinates
\be\label{cylind}
\varpi \equiv r\sin\theta, \qquad
z \equiv r\cos\theta.
\ee
In vacuum, Einstein's field equations imply that
\begin{equation}
\label{laplace}
{\partial^2 (\varpi B) \over \partial \varpi^2} 
+ {\partial^2 (\varpi B) \over \partial  z^2} =0.
\end{equation}
One can therefore define a new coordinate
\be\label{trho}
\tilde \rho \equiv \varpi B,
\ee
satisfying the two-dimensional Laplace equation (\ref{laplace}), and a
second coordinate
\begin{equation}
\tilde z = \tilde z (\varpi,z),
\end{equation}
satisfying the Cauchy-Riemann conditions
\begin{eqnarray}
\label{CauRie}
{\p \tilde z\over \p \varpi} &=& - \frac{\p \tilde\rho}{ \p z} = -\varpi \frac{\p B}{\p z}, \label{z1}\\
{\p \tilde z\over \p z}&=&{\p \tilde\rho\over \p \varpi} = B+\varpi \frac{\p B}{\p \varpi}. \label{z2}
\end{eqnarray}
The coordinate $\tilde z$ is obtained by integration of the above
Cauchy-Riemann conditions, requiring that $\tilde z=0$ in the
equatorial plane (at $z=0$).  It is easy to show that
\begin{equation}
d \varpi^2 + d  z^2 = \left[ \left( {\p \tilde \rho \over \p \varpi} \right)^2
+\left( {\p \tilde z \over \p \varpi} \right)^2  \right ]^{-1} \left ( d \tilde \rho^2 
+ d  \tilde z^2 \right),
\end{equation}
and setting
\beq
\label{MPa2}
f& = &e^{2 \nu}-\omega^2 \tilde \rho^2 e^{-2 \nu},\\
\label{MPa}
w& = &-\f{\omega \tilde \rho^2 e^{-2\nu}}
{f},\\
e^{2 \gamma} & = &  f  \left[ \left( {\p \tilde \rho \over \p \varpi} \right)^2
+\left( {\p \tilde z \over \p \varpi} \right)^2  \right ]^{-1}
 e^{2 \alpha},
\label{MPa3}
\eeq
the metric in the exterior takes the desired form (\ref{Manko}).
Since the transformation (\ref{z1},\ref{z2}) for the coordinate $\tilde z$ cannot, in
general, be solved analytically, one can relate a solution
for the interior metric (\ref{CSTmetr}) to the exterior metric
(\ref{Manko}) only through numerical integration.

\section{Tests of the accuracy of the analytic solution}
\label{checkSol}

\begin{figure}
\psfig{figure=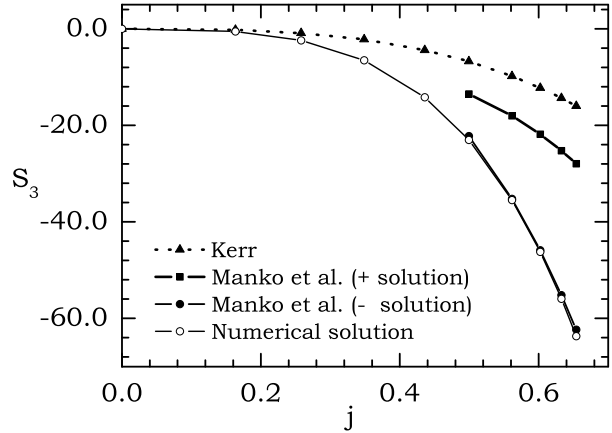,width=8cm}~~~~~ 
\caption{
The current octupole moment $S_3$ as a function of $j$ for the
numerical solution, for the Kerr metric, and for the negative and
positive branches of the analytic solution. For illustrative purposes
we have chosen the FPS EOS, and fixed our attention on the sequence
having maximum mass in the non rotating limit. The negative branch
reproduces with excellent accuracy the numerical behavior of the
current octupole, so it is the branch appropriate for describing
the exterior spacetime of rapidly rotating neutron stars.
}
\label{S3fig}
\end{figure}

\subsection{The current-octupole moment}

The current-octupole moment $S_3$, like the quadrupole moment $Q \equiv M_2$,
is a function of $a$, $b$ and $M$. Once we have fixed $b$ by matching
the quadrupole moment to the numerical spacetime through equation
(\ref{Qmatch}), there are no more free parameters to be specified; the
current-octupole $S_3$ can be computed using equation (\ref{S3}), and
then compared to the value of $S_3$ computed for the numerical
metric. Therefore, $S_3$ serves as a good error indicator for the accuracy of
the solution. In fact, the value of $S_3$ obtained analytically for
the two branches of solutions, $b_+$ and $b_-$, can be used to
distinguish which of the solutions is more relevant for rapidly
rotating neutron stars. Fig. \ref{S3fig} displays $S_3$, for the two
branches of the analytic solution, along with the value of $S_3$ for
the numerical solution and for the Kerr solution, for the evolutionary
sequence corresponding to EOS FPS and having maximum mass in the nonrotating 
limit (see Table \ref{EOSFPS}). The error for the (-)
solution is very small, at most of the order of $3~\%$. On the other
hand the error for the (+) solution is quite large (up to 56 $\%$ for
the fastest rotating model). In this case the solution is closer to
the Kerr value than to the value corresponding to numerical models of
rapidly rotating neutron stars. Therefore, in the remainder of this
paper we will only use the (-) branch of solutions to the matching
condition (\ref{Qmatch}).

In Table \ref{S3error} we display the relative error $\Delta S$ in the
analytic value of $S_3$ (when compared to the numerical solution) for
the critically rotating and maximally rotating models of all
evolutionary sequences considered in this paper. 
For most sequences, the relative error in
$S_3$ can be as large as $12\%$ for the critically rotating models,
reducing to a few percent only for the models at the mass-shedding
limit. Typically, the largest errors appear for the $1.4 M_\odot$
sequences, while the sequences that terminate at the maximum mass
static model have the smallest errors. In most cases the
error is larger for slower rotating models. This shows that for those
models $S_3$ is still influenced by its nonzero value in the
non rotating case (for the analytic solution, the octupole moment
$S_3$, like the quadrupole moment $Q$, does not vanish for $a=0$ and
$b \neq 0$). For more rapidly rotating models this influence
diminishes and $S_3$ becomes almost entirely of rotational origin,
agreeing better with the numerical solution.  Comparing the various
EOSs, one sees that the error in $S_3$ for soft EOSs, such as EOS A,
is smaller than the corresponding error for very stiff EOSs, such as
EOS L. The critical $1.4 M_\odot$ model for EOS L shows an unusually
large relative error of $45\%$ in $S_3$. This, again, is related to 
the compactness of the various models and to the value of 
the multipole  moments for $a=0$.

\subsection{Direct comparison of metric components}

\begin{figure}
\psfig{figure=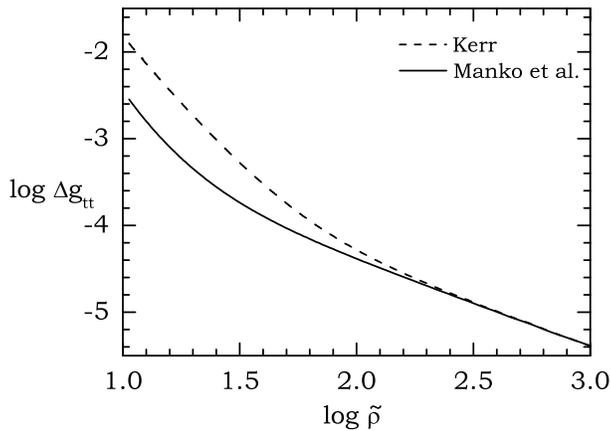,width=8cm}~~~~~ 
\caption{
Relative error in the $g_{tt}$-component of the analytic 
metric and of the Kerr metric in the equatorial plane, when
compared to the numerical solution for a rapidly rotating star.
}
\label{nu_equat}
\end{figure}

\begin{figure}
\psfig{figure=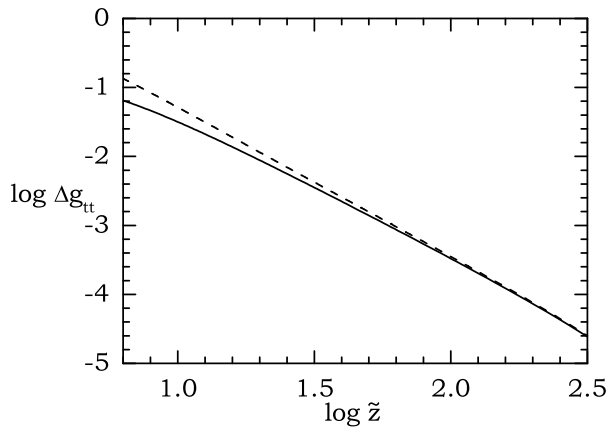,width=8cm}~~~~~ 
\caption{
Same as Fig. \ref{nu_equat}, but along the axis of symmetry.}
\label{nu_offaxis}
\end{figure}

As a second test of the accuracy of the analytic solution for rapidly
rotating neutron stars, we performed a direct comparison of specific
metric components for several representative models, using all EOSs in
our sample. Here, we focus on the most rapidly rotating model of the
maximum mass sequence with EOS FPS, since the other cases we examined
showed similar behavior.

For this model we computed the metric components $g_{tt}$, $g_{t\phi}$
and $g_{\phi\phi}$ on the equatorial plane and along the symmetry axis
using the analytic metric and the Kerr metric. Then we compared the
relative error of both metrics with respect to the corresponding
components of the numerical metric.

Fig. \ref{nu_equat} shows the relative error of the $g_{tt}$-component
of the analytic metric and of the Kerr metric in the
equatorial plane. For the analytic metric the error is only $0.3 \% $
at the surface of the star (located at $\tilde \rho=10.6$), and
decreases monotonically with increasing distance, becoming of order
$10^{-6}$ near infinity. In comparison, the relative difference
between the Kerr metric and the numerical metric is $1.3\%$ at the
equator (i.e., four times larger than the error in the analytic
metric). The relative difference between the Kerr metric and the
numerical metric also decreases with increasing distance, as expected,
and for distances larger than about 200 equatorial radii, the
difference between the analytic solution and the Kerr solution is
negligible. In other words, at such a distance the error in the analytic
solution is dominated by the Kerr contribution at first order in the
rotational parameter, while the effects of the higher-order multipole
moments $Q$ and $S_3$ have become unimportant.  The corresponding
figure for the relative error in $g_{\phi \phi}$ on the equatorial
plane is nearly identical to Fig. \ref{nu_equat} for $g_{t t}$. When
we consider $g_{t \phi}$ in the equatorial plane, the relative error
at the surface is $1.3 \%$ for the analytic metric and $5.3\%$ for the
Kerr metric. This larger error for $g_{t \phi}$ should be expected:
this metric component vanishes in the nonrotating limit, so it is more
sensitive to contributions by the higher-order multipole moments
$Q$ and $S_3$ than the metric components $g_{tt}$ and $g_{\phi \phi}$.

In order to compare the metric components on the symmetry axis, we
first need to integrate the Cauchy-Riemann conditions (\ref{CauRie})
and obtain the coordinate $\tilde z$ in terms of the coordinate
$z$. This can be done easily, once the numerical solution for the
metric function $B$ is obtained.  Fig. \ref{nu_offaxis} shows the
relative error in $g_{tt}$ for the analytic solution and the Kerr
solution along the symmetry axis. The location of the surface (as
determined from the numerical solution) is at $\tilde z=6.05$. 
At the surface, the relative error for the analytic solution is $7\%$,
while it is $15\%$ for the Kerr metric. Thus, we see that the effect
of a large quadrupole moment $Q$ shows up predominantly in the metric
components along the symmetry axis, while in the equatorial plane this
effect is very small.  The reason for this difference is that a
rapidly rotating star becomes very oblate, so that the stellar surface
on the symmetry axis is located deeper in the gravitational potential
well than the surface in the equatorial plane.  The specific example
shown in the above figures has polar to equatorial axes ratio of
$0.6$, thus the equatorial radius is roughly twice as large as the polar
radius. The analytic value of $g_{tt}$ on the surface in the
equatorial plane is $-0.59$, while it is $-0.44$ on the surface on the
symmetry axis (the asymptotic value at large distances is
$-1$). Gravity is stronger on the polar surface, and this justifies a
larger relative error in $g_{tt}$ there. At about 3 polar radii, the
relative error in $g_{tt}$ along the symmetry axis decreases to the
$1\%$ level for both the analytic and Kerr solutions.

The above direct comparison of metric components shows that the
analytic metric is a good approximation to the numerical one (or, at
least, a much better approximation than the Kerr metric) in the
equatorial plane, where one expects particle orbits to be
astrophysically more relevant. For gravitational-wave extraction in
numerical relativity, the larger inaccuracies near the polar surface
could influence the waveforms. In order to minimize this effect, the
extraction should be done as far as possible from the surface of the
star. In any case, the analytic metric is everywhere more accurate
than the Kerr metric. Thus a perturbative wave-extraction scheme,
built with the analytic metric as a background, should yield more
accurate waveforms than those obtained with techniques available at
present (which are based on a perturbative extraction of waveforms around a
Schwarzschild or Kerr background).

\subsection{Innermost stable circular orbits}

It is well known that not all orbits around relativistic stars are
stable. For nonrotating stars, the ISCO is located at a
circumferential radius of $R_{ISCO}=6M$.  Depending on the EOS and the
mass of the star, the ISCO can be located outside the stellar surface.
Rotation introduces a preferred direction in the $\phi$ coordinate, so
ISCOs around a rotating star belong to two distinct families: a
corotating and a counterrotating one. For moderate rotation rates, the
effect of rotation is to shorten the distance between the surface and
the corotating ISCO. For rapid rotation, the large quadrupole moment
of the star reverses this trend (notice that even in some Newtonian
stellar models, large higher-order multipole moments can introduce an
ISCO (see, Zdunik \& Gourgoulhon, 2001; Amsterdamski {\it et al.} 2002).
The counterrotating ISCO radius normally increases with
rotation. Detailed computations of ISCOs for a large number of models
and EOSs are presented in CST; we also refer the reader to that paper
for the equations defining the ISCOs that were used in our numerical
computations.

Testing the accuracy of the analytic solution in computing the
properties of ISCOs is important, as ISCOs are related to several
astrophysical properties of rapidly rotating neutron stars in
LMXBs. An accretion disk cannot extend to radii located within the
ISCO, and this sets an upper limit to the Keplerian frequency of
particles orbiting a star. This idea could be used, e.g., in
determining whether compact stars in LMXBs are composed of strange
matter (Stergioulas, Kluzniak \& Bulik 1999, Gondek {\it et al.}
2001). In addition, the location of the ISCO could play a role in the
mechanism producing the kHz quasi-periodic oscillations observed in
many LMXBs (van der Klis 2000): see, e.g., Kluzniak {\it et al.}
(2003).

A circular orbit in the equatorial plane is one for which $\varpi={\rm
const.}$, and hence $\tilde \rho ={\rm const.}$ The equation for
geodesic motion along the radial coordinate $\tilde \rho$ reads
\begin{eqnarray}
-g_{\tilde \rho \tilde \rho} \left(\frac{d\tilde\rho}{d\tau}\right)^2 &=&
1-\frac{{\bf E}^2 g_{\phi\phi}+2{\bf E}{\bf L} g_{t\phi}+{\bf L}^2g_{tt}}
{g^2_{t\phi}-g_{tt}g_{\phi\phi}}\equiv V(\tilde \rho),
\end{eqnarray}
where ${\bf E}$ and ${\bf L}$ are the conserved energy and angular
momentum per unit mass, determined by the conditions $V=dV/d\tilde
\rho=0$. Geodesics become unstable when $d^2V/d\tilde \rho^2=0$, or
\begin{eqnarray}
\Bigl( w'  w'' f^5\tilde\rho(2f-f' \tilde \rho)+w'^2f^4 [2f^2+(-f'^2
+f'' f)\tilde \rho^2 ] && \nn \\ 
+w' f^2 \sqrt{w'^2f^4+f' \tilde\rho(2f-f' \tilde\rho)} && \nn \\
 { [}2f^2+2f'^2\tilde\rho^2-f\tilde\rho(4f'+f'' \tilde\rho) {]}+
\tilde\rho(2f-f' \tilde\rho) && \nn \\
\bigl\{3f' f^2-4f'^2 f\tilde\rho+f'^3\tilde\rho^2+f^2[f'' \tilde\rho && \nn \\
-w'' f\sqrt{w'^2f^4+f' \tilde\rho(2f-f' \tilde\rho)} ] \bigr\} 
\Bigr) / && \nn \\
\Bigl(f^2\tilde\rho^2 \bigl\{w'^2f^4+3f' f\tilde\rho-f'^2\tilde\rho^2 && \nn \\
-f^2[2+w' \sqrt{w'^2f^4+f' \tilde\rho(2f-f' \tilde\rho)}] \bigr\} \Bigr)=0 && 
\end{eqnarray}
for corotating orbits (cf. Stute \& Camenzind, 2002), where $'$ indicates a
partial derivative with respect to $\tilde \rho$. For counter-rotating
orbits, one can simply use the above equation and change the sign of
the star's angular momentum.

Shibata \& Sasaki have used a more general representation of
axisymmetric vacuum solutions (in the form of a series expansion that
is completely determined by the physical multipole moments of the
spacetime: see Fodor, Hoenselaers and Perjes 1989 and Ryan 1995) and
derived an approximate analytic formula for the location of the
ISCO. Their formula depends on the stellar mass, angular momentum, mass
quadrupole, current octupole and mass $2^4$-pole moments. Including
all terms up to order $O(4)$ in the rotation parameter, they find the
following equation for the circumferential radius of the corotating
ISCO:
\beq
\label{SSISCO}
R_{ISCO}&=&6M(1-0.54433j-0.22619j^2+0.17989Q_2\nn\\
&-&0.23002j^3+0.26296jQ_2-0.05317q_3\nn\\
&-&0.29693j^4+0.44546j^2Q_2-0.06249Q_2^2\nn\\
&+&0.01544Q_4-0.11310jq_3).  
\eeq
In the previous expression we have introduced dimensionless parameters
$Q_2=-Q/M^3$, $q_3=-S_3/M^4$ and $Q_4=M_4/M^5$. In the case of the
Kerr metric, the approximate expression for the location of the
corotating ISCO up to order $O(j^4)$ is (see, e.g., Shapiro \&
Teukolsky 1983)
\beq
\label{kISCO}
R^{Kerr}_{ISCO}&=&6M(1-0.54433j-0.04630j^2\\
&-&0.02016j^3-0.01110j^4).\nn
\eeq
The location of the counter-rotating ISCO is obtained from the above
formulae by reversing the sign of $j$ and $S_3$.

For illustrative purposes we again focus on the three sequences of EOS
FPS (Table \ref{EOSFPS}). We have carried out the calculation for
other EOSs as well. Although there are quantitative differences
between the various models, the qualitative behavior and the relative
accuracy between the numerical and analytic solutions remain similar
to those shown here.

For each sequence we find the relative error in computing corotating
and counter-rotating ISCO radii with respect to the numerical
solution.  We perform this comparison for the analytic solution
obtained through our matching procedure (when a solution to the
matching condition exists), for the Shibata-Sasaki formula
(\ref{SSISCO}) and for the Kerr formula (\ref{kISCO}). In the case of
the Shibata-Sasaki formula we do not explicitly compute the moment
$M_4$ from the numerical solution, but we follow the same
approximation adopted by Shibata \& Sasaki. Namely, we set
$Q_4=\alpha_4 Q_2^2$, where $\alpha_4$ is expected to take values
ranging between 0 and 2. Again, following Shibata \& Sasaki, we
normally set $\alpha_4=1$ (unless otherwise noted).

\begin{figure}
\begin{center}
\leavevmode \psfig{figure=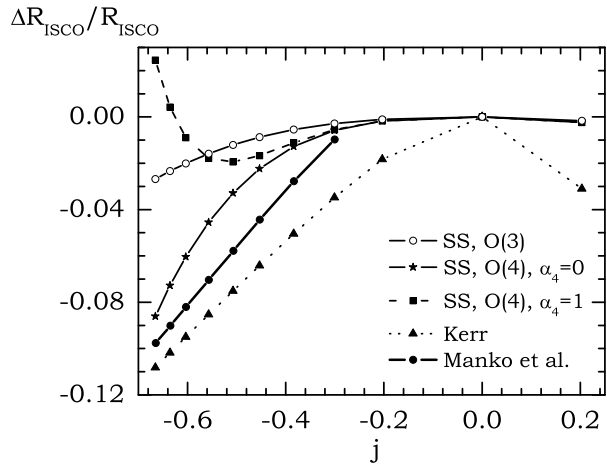,width=8cm}
\end{center}
\caption{
  Errors in the ISCO for the EOS FPS sequence with constant rest mass
  corresponding to a nonrotating model of 1.4 $M_\odot$. Negative
  values of $j$ represent counterrotating orbits. The various curves
  are explained in the main text.}
\label{14e}
\end{figure}

Fig. \ref{14e} shows the relative errors in computing the ISCOs for
the sequence having $M=1.4M_\odot$ in the non rotating limit. In this
and in the following Figures, negative values of $j$ correspond to
counterrotating orbits, while positive values of $j$ correspond to
corotating orbits. In the corotating case the ISCO disappears at slow
rotation rates, even before the analytic solution becomes valid;
therefore, for this sequence, we can only compare the accuracy in
finding counter-rotating ISCOs.  For the Kerr solution the error
increases monotonically with $|j|$, reaching $11\%$ for the fastest
rotating model.  On the other hand, the error for the Shibata-Sasaki
formula with $\alpha_4=1$ is only $~2\%$ for the fastest rotating
model. For this sequence the analytic solution is initially close to
the Shibata-Sasaki formula, but then shows a rather large error, that
becomes $~10\%$ for the fastest rotating model. The explanation for
this behavior is that at very large rotation rates the inclusion of
the multipole moment $M_4$ is important, but this multipole moment is
absent in the analytic solution. When we omit $M_4$ in the
Shibata-Sasaki formula (while still keeping all other mixed terms up
to order $O(4)$) by setting $\alpha_4=0$, we obtain an error which is
much closer to the error made using the analytic solution. On the
other hand, including only terms up to $O(3)$ in the Shibata-Sasaki
formula gives a much smaller error, comparable to (or better than) the
error of the formula when all orders up to $O(4)$ with $\alpha_4=1$
are included. What this comparison underlines is the importance of
being consistent up to a certain order in the rotation parameter.  The
Shibata-Sasaki formula has small error when used consistently up to
$O(3)$ or up to $O(4)$, but a large error when only a few mixed terms
up to $O(4)$ are included. The error of the Shibata-Sasaki formula to
order $O(4)$ should improve, if one would include the precise values
for $M_4$, instead of the crude estimate of $\alpha_4=1$.  
The analytic solution suffers from the inconsistency that while the
$M_4$ moment vanishes, it is still an exact analytic solution. This
means that mixed terms containing $j$, $Q$ and $S_3$ up to order
$O(4)$ are present. It follows that, for the counterrotating ISCOs in
Fig. \ref{14e}, the analytic solution is not as accurate as a
consistent application of the Shibata-Sasaki formula. Notice
that the non-monotonic increase in the error for the
counter-rotating ISCO with the Shibata-Sasaki formula at large
rotation rates is a consequence of the moment $M_4$ becoming important
near the mass-shedding limit: for sequences of larger mass, as the
ones we examine next, $M_4$ appears to be much less important.

\begin{figure}
\begin{center}
\leavevmode \psfig{figure=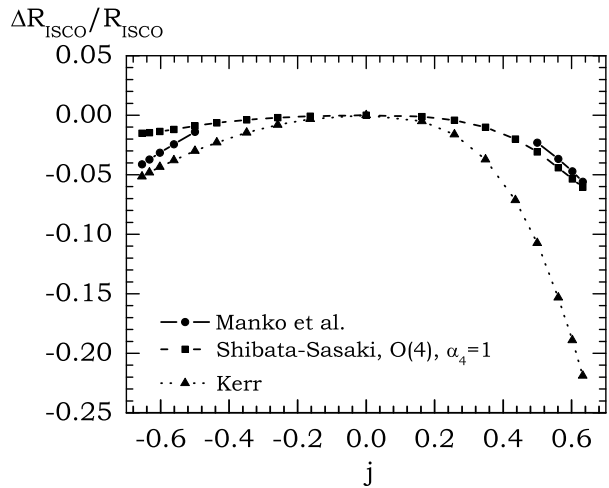,width=8cm}
\end{center}
\caption{
  Similar to Fig. \ref{14e}, for the EOS FPS sequence with constant rest mass
  corresponding to a nonrotating model of maximum mass.  }\label{MMe}
\end{figure}

\begin{figure}
\begin{center}
\leavevmode \psfig{figure=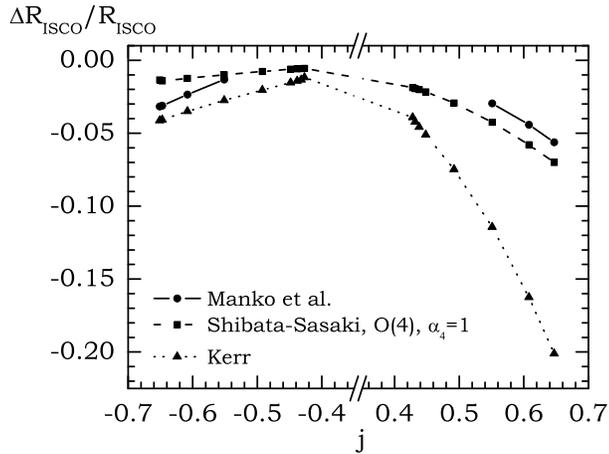,width=8cm}
\end{center}
\caption{
  Similar to Fig. \ref{14e}, for a supramassive EOS FPS sequence with
  constant rest mass. }\label{SMe}
\end{figure}

The comparison of the error in computing the ISCOs is much more
favorable for the analytic solution in the case of the other two
sequences we examined.  Fig. \ref{MMe} shows the errors for the
evolutionary sequence that terminates at the maximum-mass nonrotating
model. In this case, a corotating ISCO exists for some models for
which the analytic solution is valid. The error made with the analytic
solution is $~5\%$ for the fastest rotating model, and it is
consistently better than the error of the Shibata-Sasaki formula. For
counter-rotating orbits the error for the analytic solution is
somewhat smaller than for corotating orbits (which is expected, as the
ISCO for corotating orbits is normally at larger radii). However, the
error is consistently larger than the error of the Shibata-Sasaki
formula. The corresponding errors for the supramassive sequence, shown
in Fig. \ref{SMe}, are very similar to the errors for the sequence in
Fig. \ref{MMe}.


\subsection{Comparison to other matching conditions}

Manko {\it et al.} (2000a) also use the quadrupole moment $Q$ in order
to match the analytic solution to a numerical one. However, they
redefine the parameter $b$ as
\begin{equation}
b=\pm \sqrt{a^2+2aM\Delta-M^2},
\end{equation}
where $\Delta$ is a new parameter, with the motivation that now
$\Delta$ measures the departure of the analytic solution from the Kerr
metric.  We find that the above redefinition is not necessary, as it
does not change the solution: in other words, a solution with a given
$b$ has a corresponding value of $\Delta$.  In Manko {\it et al.}
(2000a) it is not mentioned that $Q$ can be set to the numerical value
only for a limited range of the parameter $b$ (or, equivalently, of
$\Delta$). Moreover, we find that the illustrative solutions they give
do {\it not} correspond to the negative branch of solutions for $b$
(that, as we have seen, are those relevant for rapidly rotating stars)
but rather to the positive branch, which is closer in behavior to the
Kerr metric.

Stute \& Camenzind (2002) fix the the third parameter in the analytic
solution, $b$, by matching the value of the metric function $g_{tt}$
at the stellar equator. However, this is a local quantity, so that the
analytic and numerical metrics are matched only at a single point in
the $(\tilde \rho, \tilde z)$ plane. Experimenting with this choice,
we found that one can obtain numerical solutions that are also limited
to rapidly rotating stars. However, since the parameter $b$ is not
fixed directly, but only indirectly, one has to solve a nonlinear
equation in order to obtain $b$ for a given value of $g_{tt}$ at the
equator. This procedure could lead to multiple solutions, and one has
to choose the one closest to a rapidly rotating neutron star by
examining other properties of the spacetime (e.g. the higher multipole
moments). As we have seen, fixing $Q$ also leads to multiple solutions,
however, we find that the procedure of fixing directly three leading
multipole moments ($M$, $j$ and $Q$) and selecting the desired
solution according to the value of a fourth multipole moment ($S_3$)
is more intuitive that fixing $M$, $j$ and the value of a metric
function a single point in the spacetime.

In the matching procedure used by Stute \& Camenzind the parameter $b$
is not chosen to be real. They rather impose that $b$ continuously
reduces to the Schwarzschild value, $b=i M$, in the nonrotating limit
($a\to 0$). However, the Schwarzschild and Kerr solutions can only
(formally) be obtained as limiting cases of the Manko solution by
analytic continuation in the complex-$b$ plane. In a sense, the black
hole solutions are ``isolated points'' on the pure-imaginary axis of
the complex-$b$ plane, while solutions representing neutron star
exteriors lie on the pure-real axis.  Therefore, the requirement
imposed by Stute \& Camenzind violates one of the original
requirements of the analytic solution (namely, that all three
parameters of the solution must be real).  If one follows this
procedure, the resulting metric components are, in general, complex.
For complex values of $b$ one could, in principle, use the real parts
of some quantities in order to compute an estimate for the location of
the ISCO, as was done by Stute \& Camenzind. However, in such cases,
even the coordinates in which the metric is expressed become complex
numbers. Furthermore, additional multipole moments appear that are not
present in the numerical solution, rendering the analytic solution
inappropriate for describing the physical properties of a rotating
neutron star. 

Finally, an important point in the matching procedure is to use the
correct correspondence between the coordinates in the analytic
exterior spacetime (\ref{Manko}) and the numerical spacetime
(\ref{CSTmetr}). Stute \& Camenzind transformed the analytic metric to
Boyer-Lindquist like coordinates, but these are {\it not} the
coordinates used in (\ref{CSTmetr}). This can easily be seen when one
considers that the metric (\ref{CSTmetr}) reduces to the Schwarzschild
metric in {\it isotropic} coordinates (not in the usual Schwarzschild
coordinates) in the nonrotating limit.

\section{Conclusions}

We have investigated the properties of a closed-form analytic solution
for the exterior spacetime of rapidly rotating neutron stars.  We
matched it to highly-accurate numerical solutions, imposing that the
quadrupole moment of the numerical and analytic spacetimes be the
same. For the analytic solution we considered, such a matching
condition can be satisfied only for very rapidly rotating stars. We
found that solutions belong to two branches, only one of which is a
good approximation to the exterior of rapidly rotating neutron star
spacetimes. In order to evaluate the accuracy of the analytic solution
in describing rapidly rotating neutron stars, we presented a
comparison of the radii of ISCOs obtained with a) the analytic
solution, b) the Kerr metric, c) an analytic series-expansion derived
by Shibata \& Sasaki and d) a highly-accurate numerical code. In most
cases we found that the analytic solution has an accuracy consistently
better than the Shibata-Sasaki expansion up to $O(j^4)$, for
corotating orbits. Only for counterrotating orbits does the
higher-order Shibata-Sasaki expansion perform better than the analytic
solution. We have only shown direct comparisons for three constant
rest-mass sequences and one representative EOS (FPS); however our
qualitative conclusions also hold for other EOSs.

The analytic solution we studied in this paper could become useful in
constructing outgoing-wave boundary conditions for simulations of
pulsating relativistic stars, and for the computation of quasinormal
modes of oscillation as an eigenvalue problem (a long-standing problem
in relativistic astrophysics). Another potential application is the
study of high-frequency variability in accretion disks around rapidly
rotating relativistic stars.  We emphasize, however, that this
analytic solution is only valid for rapidly rotating stars, contrary
to previous claims in the literature. For stars of intermediate
rotation rates one can use the exterior analytic solution by Hartle \&
Thorne (1968), valid to second order in the rotation rate.  This
approximate solution is determined by the three multipole moments $M$,
$j$ and $Q$, but higher-order multipole moments are ignored. It would
be interesting to determine whether the region in which the
second-order Hartle-Thorne metric is valid to some accuracy, overlaps
with the region in which the analytic solution considered here is
valid. Such a study, along with a characterization of the spacetimes
using invariant quantities (constructed in the Newman-Penrose
formalism) will be reported elsewhere (Berti {\it et al.}, in
preparation).

\begin{center}
\bf{Acknowledgements}
\end{center}
We wish to thank Marco Bruni, John L. Friedman, Kostas Kokkotas, Mina
Maniopoulou, Vladimir Manko, Masaru Shibata, Nail Sibgatullin, Frances
White and Leszek Zdunik for useful discussions and correspondence. We
are grateful to the anonymous referee for a very careful reading of the
paper and many suggestions. This work has been supported by the EU
Programme 'Improving the Human Research Potential and the
Socio-Economic Knowledge Base' (Research Training Network Contract
HPRN-CT-2000-00137).


\clearpage

\begin{table}
\begin{center}
\begin{tabular}{||p{0.01cm}*{12}{c|}|}
\hline
&$\epsilon_c$ &$\Omega$ &$I$ &$M$ &$T/W$ &$R_e$ &$h_+$ &$h_-$ &$M_2$ &$J$ &$S_3$ &$b$\\
&$10^{15}$~g~cm$^{-3}$ &$10^3$~s$^{-1}$ &$10^{45}$g~cm$^2$ &$M_\odot$ &$-$ &km &km &km &km$^3$ &km$^2$ &km$^4$ &km\\
\hline
\hline
\multicolumn{13}{c}{$M_B=1.589 M_\odot$}\\
\hline 
&$1.8582$ &$0.000$ &$-$ &$1.402$ &$0.0000$ &$9.570$ &$2.841$ &$2.841$ &$0.000$   &$0.0000$ &$0.000$  &$-$  \\
&$1.8127$ &$3.205$ &$1.023$ &$1.405$ &$0.0099$ &$9.741$ &$1.609$ &$4.115$ &$-1.001$  &$0.8121$ &$-0.727$ &$-$  \\
&$1.7422$ &$5.005$ &$1.070$ &$1.410$ &$0.0257$ &$10.04$ &$0.924$ &$4.866$ &$-2.656$  &$1.327$  &$-3.158$ &$-$  \\
&$1.6744$ &$6.138$ &$1.121$ &$1.415$ &$0.0411$ &$10.36$ &$0.485$ &$5.354$ &$-4.377$  &$1.704$  &$-6.694$ &$-0.4777$  \\
&$1.6093$ &$6.952$ &$1.174$ &$1.419$ &$0.0563$ &$10.71$ &$0.151$ &$5.706$ &$-6.173$  &$2.022$  &$-11.23$ &$-0.6302$  \\
&$1.5467$ &$7.565$ &$1.232$ &$1.424$ &$0.0712$ &$11.12$ &$-$     &$5.951$ &$-8.063$  &$2.307$  &$-16.77$ &$-0.6738$  \\
&$1.4939$ &$7.982$ &$1.285$ &$1.428$ &$0.0840$ &$11.53$ &$-$     &$6.073$ &$-9.806$  &$2.540$  &$-22.50$ &$-0.6701$  \\
&$1.4502$ &$8.266$ &$1.333$ &$1.432$ &$0.0948$ &$11.97$ &$-$     &$6.083$ &$-11.37$  &$2.729$  &$-28.10$ &$-0.6486$  \\
&$1.4146$ &$8.463$ &$1.376$ &$1.435$ &$0.1038$ &$12.46$ &$-$     &$5.961$ &$-12.74$  &$2.884$  &$-33.35$ &$-0.6214$  \\
&$1.4050$ &$8.511$ &$1.388$ &$1.435$ &$0.1062$ &$12.64$ &$-$     &$5.883$ &$-13.13$  &$2.925$  &$-34.88$ &$-0.6128$  \\
\hline
\multicolumn{13}{c}{$M_B=1.948 M_\odot$}\\
\hline 
&$4.1300$ &$0.000$ &$-$  &$1.658$ &$0.0000$  &$8.362$  &$6.313$ &$6.313$ &$0.000$ &$0.0000$  &$0.000$   &$-$\\
&$3.4156$ &$3.571$ &$1.115$  &$1.662$ &$0.0734$  &$8.740$  &$4.701$ &$7.315$ &$-0.623$ &$0.9857$ &$-0.376$  &$-$\\
&$3.0643$ &$5.410$ &$1.174$  &$1.668$ &$0.0186$  &$9.042$  &$3.776$ &$7.897$ &$-1.657$ &$1.573$  &$-1.637$  &$-$\\
&$2.7492$ &$6.942$ &$1.249$  &$1.676$ &$0.0341$  &$9.419$  &$2.911$ &$8.422$ &$-3.237$ &$2.147$  &$-4.459$  &$-$\\
&$2.4664$ &$8.172$ &$1.343$  &$1.686$ &$0.0533$  &$9.894$  &$2.104$ &$8.883$ &$-5.434$ &$2.717$  &$-9.667$  &$-$\\
&$2.2736$ &$8.893$ &$1.427$  &$1.694$ &$0.0696$  &$10.34$  &$1.540$ &$9.163$ &$-7.529$ &$3.143$  &$-15.72$  &$-0.2679$\\
&$2.1245$ &$9.371$ &$1.509$  &$1.701$ &$0.0842$  &$10.79$  &$1.081$ &$9.336$ &$-9.619$ &$3.501$  &$-22.64$  &$-0.4798$\\
&$2.0123$ &$9.595$ &$1.576$  &$1.708$ &$0.0945$  &$11.20$  &$0.723$ &$9.356$ &$-11.31$ &$3.746$  &$-28.76$  &$-0.5568$\\
&$1.9060$ &$9.852$ &$1.660$  &$1.714$ &$0.1078$  &$11.79$  &$0.252$ &$9.309$ &$-13.52$ &$4.051$  &$-37.53$  &$-0.5941$\\
&$1.8500$ &$10.00$ &$1.715$  &$1.719$ &$0.1163$  &$12.38$  &$-$     &$9.092$ &$-15.03$ &$4.247$  &$-44.00$  &$-0.5972$\\
\hline
\multicolumn{13}{c}{$M_B=2.038 M_\odot$}\\
\hline 
&$3.8786$ &$8.671$ &$1.210$ &$1.742$ &$0.0416$ &$8.892$  &$3.443$ &$9.917$  &$-3.489$ &$2.598$ &$-4.943$   &$-$\\
&$3.7154$ &$8.603$ &$1.227$ &$1.742$ &$0.0423$ &$8.977$  &$3.363$ &$9.867$  &$-3.615$ &$2.614$ &$-5.221$   &$-$\\
&$3.5592$ &$8.597$ &$1.246$ &$1.743$ &$0.0436$ &$9.071$  &$3.255$ &$9.840$  &$-3.809$ &$2.653$ &$-5.654$   &$-$\\
&$3.4095$ &$8.646$ &$1.267$ &$1.744$ &$0.0457$ &$9.175$  &$3.123$ &$9.836$  &$-4.077$ &$2.714$ &$-6.266$   &$-$\\
&$3.2660$ &$8.742$ &$1.291$ &$1.745$ &$0.0485$ &$9.291$  &$2.967$ &$9.853$  &$-4.423$ &$2.796$ &$-7.082$   &$-$\\
&$2.9334$ &$9.123$ &$1.361$ &$1.751$ &$0.0583$ &$9.633$  &$2.499$ &$9.963$  &$-5.653$ &$3.076$ &$-10.24$  &$-$\\
&$2.4701$ &$9.866$ &$1.513$ &$1.765$ &$0.0817$ &$10.40$ &$1.593$ &$10.22$ &$-8.890$ &$3.697$ &$-20.13$   &$-0.2083$\\
&$2.2667$ &$10.19$ &$1.611$ &$1.773$ &$0.0964$ &$10.94$ &$1.077$ &$10.29$ &$-11.21$ &$4.064$ &$-28.43$   &$-0.4423$\\
&$2.1252$ &$10.37$ &$1.697$ &$1.780$ &$0.1086$ &$11.51$ &$0.606$ &$10.24$ &$-13.32$ &$4.361$ &$-36.78$   &$-0.5274$\\
&$2.0800$ &$10.42$ &$1.728$ &$1.782$ &$0.1128$ &$11.76$ &$0.401$ &$10.16$ &$-14.09$ &$4.462$ &$-40.03$   &$-0.5443$\\
\hline
\end{tabular}
\end{center}
\caption{Equilibrium properties for three sequences of constant rest mass $M_B$, constructed with EOS A. We show a sequence that corresponds to a gravitational mass of $M=1.4M_\odot$ in the nonrotating limit ($M_B=1.589 M_\odot$), a sequence that terminates at the maximum-mass nonrotating model in the nonrotating limit ($M_B=1.948 M_\odot$) and a supramassive sequence ($M_B=2.038 M_\odot$).}\label{EOSA}
\end{table}

\clearpage

\begin{table}
\begin{center}
\begin{tabular}{||p{0.01cm}*{12}{c|}|}
\hline
&$\epsilon_c$ &$\Omega$ &$I$ &$M$ &$T/W$ &$R_e$ &$h_+$ &$h_-$ &$M_2$ &$J$ &$S_3$ &$b$\\
&$10^{15}$~g~cm$^{-3}$ &$10^3$~s$^{-1}$ &$10^{45}$g~cm$^2$ &$M_\odot$ &$-$ &km &km &km &km$^3$ &km$^2$ &km$^4$ &km\\
\hline
\hline
\multicolumn{13}{c}{$M_B=1.578 M_\odot$}\\
\hline 
&$1.2062$ &$0.000$ &$-$ &$1.402$ &$0.0000$ &$10.39$ &$2.020$ &$2.020$ &$0.000$ &$0.0000$ &$0.000$     &$-$   \\
&$1.1922$ &$2.951$ &$1.191$ &$1.405$ &$0.0111$ &$10.57$ &$0.796$ &$3.452$ &$-1.416$ &$0.8704$ &$-1.126$   &$-$   \\
&$1.1728$ &$4.431$ &$1.235$ &$1.409$ &$0.0262$ &$10.84$ &$0.277$ &$4.228$ &$-3.364$ &$1.355$ &$-4.168$   &$-0.3280$   \\
&$1.1483$ &$5.618$ &$1.292$ &$1.415$ &$0.0446$ &$11.20$ &$-$     &$4.867$ &$-5.818$ &$1.798$ &$-9.564$   &$-0.6683$   \\
&$1.1296$ &$6.276$ &$1.339$ &$1.418$ &$0.0582$ &$11.51$ &$-$     &$5.214$ &$-7.703$ &$2.081$ &$-14.66$  &$-0.7247$   \\
&$1.1112$ &$6.789$ &$1.386$ &$1.422$ &$0.0712$ &$11.85$ &$-$     &$5.462$ &$-9.581$ &$2.331$ &$-20.44$  &$-0.7266$   \\
&$1.0931$ &$7.196$ &$1.435$ &$1.426$ &$0.0836$ &$12.24$ &$-$     &$5.615$ &$-11.46$ &$2.558$ &$-26.87$ &$-0.7019$   \\
&$1.0779$ &$7.478$ &$1.479$ &$1.429$ &$0.0938$ &$12.63$ &$-$     &$5.655$ &$-13.08$ &$2.739$ &$-32.87$ &$-0.6691$   \\
&$1.0628$ &$7.708$ &$1.522$ &$1.431$ &$0.1035$ &$13.13$ &$-$     &$5.562$ &$-14.68$ &$2.906$ &$-39.20$ &$-0.6305$   \\
&$1.0580$ &$7.772$ &$1.536$ &$1.432$ &$0.1065$ &$13.34$ &$-$     &$5.479$ &$-15.19$ &$2.957$ &$-41.32$ &$-0.6173$   \\
\hline
\multicolumn{13}{c}{$M_B=2.636 M_\odot$}\\
\hline
&$3.0200$ &$0.000$ &$-$ &$2.136$ &$0.0000$ &$9.405$ &$9.507$ &$9.507$  &$0.000$ &$0.000$ &$0.000$    &$-$\\
&$2.5479$ &$3.765$ &$2.080$ &$2.145$ &$0.0098$ &$9.739$ &$7.269$ &$11.26$ &$-1.370$ &$1.939$ &$-1.031$   &$-$\\
&$2.3008$ &$5.831$ &$2.173$ &$2.158$ &$0.0256$ &$10.04$ &$5.935$ &$12.32$ &$-3.701$ &$3.138$ &$-4.622$   &$-$\\
&$2.1014$ &$7.384$ &$2.283$ &$2.174$ &$0.0447$ &$10.39$ &$4.822$ &$13.18$ &$-6.740$ &$4.176$ &$-11.44$  &$-$\\
&$1.9411$ &$8.503$ &$2.402$ &$2.190$ &$0.0645$ &$10.78$ &$3.916$ &$13.85$ &$-10.14$ &$5.059$ &$-21.24$ &$-$\\
&$1.8135$ &$9.298$ &$2.526$ &$2.205$ &$0.0836$ &$11.19$ &$3.166$ &$14.36$ &$-13.72$ &$5.817$ &$-33.57$ &$-$\\
&$1.7136$ &$9.847$ &$2.651$ &$2.219$ &$0.1010$ &$11.62$ &$2.541$ &$14.71$ &$-17.30$ &$6.465$ &$-47.71$ &$-$\\
&$1.6377$ &$10.20$ &$2.766$ &$2.231$ &$0.1156$ &$12.07$ &$2.016$ &$14.91$ &$-20.58$ &$6.986$ &$-62.16$ &$-0.1302$\\
&$1.5829$ &$10.41$ &$2.865$ &$2.240$ &$0.1271$ &$12.53$ &$1.553$ &$14.96$ &$-23.37$ &$7.389$ &$-75.49$ &$-0.3222$\\
&$1.5327$ &$10.57$ &$2.970$ &$2.249$ &$0.1382$ &$13.34$ &$0.793$ &$14.64$ &$-26.32$ &$7.778$ &$-90.56$ &$-0.4275$\\
\hline
\multicolumn{13}{c}{$M_B=2.799 M_\odot$}\\
\hline
&$2.7800$ &$9.849$ &$2.401$ &$2.297$ &$0.0682$ &$10.16$ &$4.647$ &$15.80$ &$-10.16$ &$5.856$ &$-19.56$  &$-$\\
&$2.7266$ &$9.815$ &$2.413$ &$2.297$ &$0.0686$ &$10.20$ &$4.614$ &$15.77$ &$-10.28$ &$5.865$ &$-19.95$  &$-$\\
&$2.6250$ &$9.783$ &$2.439$ &$2.298$ &$0.0699$ &$10.28$ &$4.526$ &$15.75$ &$-10.59$ &$5.909$ &$-20.99$  &$-$\\
&$2.5271$ &$9.802$ &$2.470$ &$2.300$ &$0.0721$ &$10.38$ &$4.403$ &$15.76$ &$-11.06$ &$5.995$ &$-22.55$  &$-$\\
&$2.2982$ &$9.991$ &$2.562$ &$2.308$ &$0.0807$ &$10.67$ &$3.979$ &$15.87$ &$-12.85$ &$6.339$ &$-28.61$  &$-$\\
&$2.0900$ &$10.35$ &$2.688$ &$2.320$ &$0.0946$ &$11.07$ &$3.390$ &$16.12$ &$-15.74$ &$6.886$ &$-39.28$  &$-$\\
&$1.9372$ &$10.66$ &$2.814$ &$2.333$ &$0.1085$ &$11.48$ &$2.836$ &$16.33$ &$-18.87$ &$7.427$ &$-52.04$  &$-$\\
&$1.8299$ &$10.88$ &$2.928$ &$2.344$ &$0.1207$ &$11.90$ &$2.351$ &$16.46$ &$-21.80$ &$7.888$ &$-65.12$  &$-$\\
&$1.7285$ &$11.06$ &$3.066$ &$2.357$ &$0.1343$ &$12.51$ &$1.721$ &$16.46$ &$-25.39$ &$8.400$ &$-82.46$  &$0.0939$\\
&$1.6800$ &$11.14$ &$3.145$ &$2.364$ &$0.1416$ &$13.15$ &$1.108$ &$16.15$ &$-27.46$ &$8.674$ &$-93.17$  &$-0.1280$\\
\hline
\end{tabular}
\end{center}
\caption{Same as Table \ref{EOSA}, but for EOS AU.}
\label{EOSAU}
\end{table}

\clearpage

\begin{table}
\begin{center}
\begin{tabular}{||p{0.01cm}*{12}{c|}|}
\hline
&$\epsilon_c$ &$\Omega$ &$I$ &$M$ &$T/W$ &$R_e$ &$h_+$ &$h_-$ &$M_2$ &$J$ &$S_3$ &$b$\\
&$10^{15}$~g~cm$^{-3}$ &$10^3$~s$^{-1}$ &$10^{45}$g~cm$^2$ &$M_\odot$ &$-$ &km &km &km &km$^3$ &km$^2$ &km$^4$ &km\\
\hline
\multicolumn{13}{c}{$M_B=1.561 M_\odot$}\\
\hline
&$1.2974$& $0.000$& $-$& $1.402$& $0.0000$& $10.85$& $1.560$& $1.560$& $0.000$&   $0.000$& $0.000$   &$-$\\
&$1.2660$& $2.844$& $1.238$& $1.404$& $0.0111$& $11.07$& $0.312$& $2.970$& $-1.512$&  $0.872$& $-1.241$  &$-$\\
&$1.2303$& $4.076$& $1.285$& $1.408$& $0.0238$& $11.35$& $-$    & $3.622$& $-3.312$&  $1.297$& $-4.051$  &$-0.3392$\\
&$1.1908$& $5.012$& $1.341$& $1.412$& $0.0382$& $11.70$& $-$    & $4.129$& $-5.417$&  $1.665$& $-8.518$  &$-0.6669$\\
&$1.1525$& $5.696$& $1.401$& $1.415$& $0.0523$& $12.09$& $-$    & $4.493$& $-7.598$&  $1.977$& $-14.21$ &$-0.7508$\\
&$1.1201$& $6.159$& $1.456$& $1.419$& $0.0643$& $12.48$& $-$    & $4.715$& $-9.573$&  $2.222$& $-20.16$ &$-0.7596$\\
&$1.0885$& $6.532$& $1.515$& $1.422$& $0.0763$& $12.92$& $-$    & $4.846$& $-11.64$& $2.451$& $-27.10$ &$-0.7381$\\
&$1.0578$& $6.831$& $1.577$& $1.425$& $0.0881$& $13.48$& $-$    & $4.847$& $-13.79$& $2.668$& $-35.06$ &$-0.6986$\\
&$1.0364$& $7.007$& $1.624$& $1.428$& $0.0964$& $14.02$& $-$    & $4.699$& $-15.40$& $2.819$& $-41.46$ &$-0.6629$\\
&$1.0157$& $7.150$& $1.672$& $1.430$& $0.1044$& $15.05$& $-$    & $4.049$& $-17.02$& $2.961$& $-48.26$ &$-0.6244$\\
\hline
\multicolumn{13}{c}{$M_B=2.105 M_\odot$}\\
\hline
&$3.3900$&  $0.000$& $-$& $1.802$& $0.0000$& $9.276$& $6.674$& $6.674$&  $0.000$&   $0.000$& $0.000$   &$-$\\
&$2.7939$&  $3.204$& $1.462$& $1.806$& $0.0073$& $9.708$& $4.911$& $7.741$&  $-0.830$&  $1.160$& $-0.556$  &$-$\\
&$2.5016$&  $4.833$& $1.543$& $1.813$& $0.0184$& $10.05$& $3.904$& $8.350$& $-2.200$&   $1.847$& $-2.402$  &$-$\\
&$2.2399$&  $6.184$& $1.645$& $1.821$& $0.0337$& $10.49$& $2.961$& $8.900$& $-4.303$&   $2.519$& $-6.549$  &$-$\\
&$2.0056$&  $7.260$& $1.772$& $1.831$& $0.0525$& $11.03$& $2.087$& $9.374$& $-7.222$&   $3.185$& $-14.18$ &$-$\\
&$1.8461$&  $7.887$& $1.885$& $1.840$& $0.0685$& $11.53$& $1.477$& $9.659$& $-10.01$&   $3.683$& $-23.05$ &$-0.3792$\\
&$1.6992$&  $8.370$& $2.017$& $1.849$& $0.0856$& $12.16$& $0.881$& $9.846$& $-13.37$&   $4.181$& $-35.48$ &$-0.5965$\\
&$1.6079$&  $8.613$& $2.117$& $1.855$& $0.0976$& $12.70$& $0.449$& $9.867$& $-15.97$&   $4.515$& $-46.22$ &$-0.6566$\\
&$1.5426$&  $8.760$& $2.199$& $1.860$& $0.1069$& $13.24$& $0.035$& $9.759$& $-18.16$&   $4.771$& $-55.96$ &$-0.6736$\\
&$1.5000$&  $8.846$& $2.260$& $1.864$& $0.1135$& $13.82$& $-$    & $9.492$& $-19.80$&   $4.951$& $-63.67$ &$-0.6739$\\
\hline
\multicolumn{13}{c}{$M_B=2.226 M_\odot$}\\
\hline
&$3.2103$ &$8.452$ &$1.628$ &$1.914$ &$0.0492$ &$9.977$ &$3.346$ &$11.04$ &$-5.570$ &$3.409$ &$-9.594$   &$-$\\
&$3.0701$ &$8.350$ &$1.652$ &$1.914$ &$0.0496$ &$10.07$ &$3.268$ &$10.96$ &$-5.733$ &$3.417$ &$-10.03$  &$-$\\
&$2.9361$ &$8.301$ &$1.679$ &$1.915$ &$0.0507$ &$10.18$ &$3.163$ &$10.92$ &$-5.991$ &$3.452$ &$-10.72$  &$-$\\
&$2.8079$ &$8.284$ &$1.709$ &$1.916$ &$0.0524$ &$10.30$ &$3.035$ &$10.88$ &$-6.328$ &$3.506$ &$-11.64$  &$-$\\
&$2.6854$ &$8.326$ &$1.742$ &$1.918$ &$0.0550$ &$10.44$ &$2.875$ &$10.88$ &$-6.802$ &$3.593$ &$-12.98$  &$-$\\
&$2.3488$ &$8.607$ &$1.864$ &$1.925$ &$0.0665$ &$10.94$ &$2.279$ &$10.96$ &$-8.917$ &$3.973$ &$-19.46$  &$-$\\
&$2.0544$ &$8.973$ &$2.024$ &$1.936$ &$0.0833$ &$11.62$ &$1.554$ &$11.08$ &$-12.24$ &$4.498$ &$-31.21$ &$-0.3080$\\
&$1.8375$ &$9.239$ &$2.193$ &$1.946$ &$0.1007$ &$12.45$ &$0.831$ &$11.07$ &$-16.13$ &$5.018$ &$-47.06$ &$-0.5536$\\
&$1.7185$ &$9.377$ &$2.318$ &$1.954$ &$0.1131$ &$13.27$ &$0.162$ &$10.83$ &$-19.18$ &$5.383$ &$-60.96$ &$-0.6168$\\
&$1.6800$ &$9.333$ &$2.346$ &$1.955$ &$0.1145$ &$13.56$ &$-$     &$10.60$ &$-19.76$ &$5.423$ &$-63.67$ &$-0.6293$\\
\hline
\end{tabular}
\end{center}
\caption{Same as Table \ref{EOSA}, but for EOS FPS.}
\label{EOSFPS}
\end{table}

\clearpage

\begin{table}
\begin{center}
\begin{tabular}{||p{0.01cm}*{12}{c|}|}
\hline
&$\epsilon_c$ &$\Omega$ &$I$ &$M$ &$T/W$ &$R_e$ &$h_+$ &$h_-$ &$M_2$ &$J$ &$S_3$ &$b$\\
&$10^{15}$~g~cm$^{-3}$ &$10^3$~s$^{-1}$ &$10^{45}$g~cm$^2$ &$M_\odot$ &$-$ &km &km &km &km$^3$ &km$^2$ &km$^4$ &km\\
\hline
\hline
\multicolumn{13}{c}{$M_B=1.510 M_\odot$}\\
\hline
&$0.4326$ &$0.000$ &$-$ &$1.402$ &$0.0000$ &$14.83$ &$-$     &$-$     &$0.000$ &$0.0000$ &$0.000$      &$-$\\
&$0.4266$ &$1.816$ &$2.162$ &$1.404$ &$0.0124$ &$15.14$ &$-$     &$-$     &$-3.671$ &$0.9726$ &$-3.554$    &$-0.5958$\\
&$0.4188$ &$2.690$ &$2.253$ &$1.407$ &$0.0286$ &$15.58$ &$-$     &$0.541$ &$-8.464$ &$1.501$ &$-12.65$   &$-0.9422$\\
&$0.4111$ &$3.268$ &$2.346$ &$1.410$ &$0.0443$ &$16.06$ &$-$     &$1.283$ &$-13.20$ &$1.899$ &$-24.96$  &$-0.9455$\\
&$0.4045$ &$3.648$ &$2.431$ &$1.412$ &$0.0577$ &$16.52$ &$-$     &$1.749$ &$-17.29$ &$2.196$ &$-37.84$  &$-0.8901$\\
&$0.3980$ &$3.954$ &$2.519$ &$1.415$ &$0.0709$ &$17.05$ &$-$     &$2.077$ &$-21.42$ &$2.466$ &$-52.65$  &$-0.8151$\\
&$0.3916$ &$4.204$ &$2.611$ &$1.417$ &$0.0839$ &$17.68$ &$-$     &$2.250$ &$-25.60$ &$2.719$ &$-69.47$  &$-0.7311$\\
&$0.3853$ &$4.407$ &$2.707$ &$1.420$ &$0.0965$ &$18.49$ &$-$     &$2.195$ &$-29.80$ &$2.955$ &$-88.00$  &$-0.6444$\\
&$0.3800$ &$4.552$ &$2.793$ &$1.421$ &$0.1070$ &$19.61$ &$-$     &$1.693$ &$-33.46$ &$3.148$ &$-105.4$ &$-0.5693$\\
&$0.3790$ &$4.573$ &$2.807$ &$1.422$ &$0.1087$ &$19.97$ &$-$     &$1.430$ &$-34.06$ &$3.179$ &$-108.4$ &$-0.5569$\\
\hline
\multicolumn{13}{c}{$M_B=3.232 M_\odot$}\\
\hline
&$1.4700$ &$0.000$ &$-$ &$2.713$ &$0.0000$ &$13.71$ &$10.31$ &$10.31$ &$0.000$ &$0.0000$ &$0.000$       &$-$\\
&$1.2010$ &$2.063$ &$5.009$ &$2.720$ &$0.0069$ &$14.20$ &$7.868$ &$12.01$ &$-2.656$ &$2.559$ &$-2.558$     &$-$\\
&$1.0639$ &$3.339$ &$5.273$ &$2.731$ &$0.0200$ &$14.65$ &$6.260$ &$13.22$ &$-8.015$ &$4.361$ &$-13.43$   &$-$\\
&$0.9552$ &$4.343$ &$5.591$ &$2.746$ &$0.0372$ &$15.19$ &$4.888$ &$14.27$ &$-15.78$ &$6.014$ &$-37.00$   &$-$\\
&$0.8692$ &$5.088$ &$5.948$ &$2.762$ &$0.0564$ &$15.80$ &$3.767$ &$15.14$ &$-25.24$ &$7.496$ &$-74.63$   &$-$\\
&$0.8017$ &$5.608$ &$6.320$ &$2.778$ &$0.0752$ &$16.45$ &$2.889$ &$15.80$ &$-35.49$ &$8.777$ &$-124.1$  &$-0.5411$\\
&$0.7495$ &$5.964$ &$6.692$ &$2.793$ &$0.0927$ &$17.14$ &$2.189$ &$16.28$ &$-45.99$ &$9.885$ &$-182.6$  &$-0.8287$\\
&$0.7101$ &$6.192$ &$7.037$ &$2.806$ &$0.1077$ &$17.82$ &$1.623$ &$16.56$ &$-55.82$ &$10.79$ &$-243.8$ &$-0.9222$\\
&$0.6729$ &$6.371$ &$7.431$ &$2.820$ &$0.1234$ &$18.73$ &$0.958$ &$16.65$ &$-67.15$ &$11.73$ &$-321.3$ &$-0.9373$\\
&$0.6600$ &$6.423$ &$7.586$ &$2.825$ &$0.1293$ &$19.19$ &$0.628$ &$16.57$ &$-71.63$ &$12.07$ &$-353.8$ &$-0.9275$\\
\hline
\multicolumn{13}{c}{$M_B=3.470 M_\odot$}\\
\hline
&$1.3847$ &$6.095$ &$5.855$ &$2.929$ &$0.0607$ &$14.86$ &$5.000$ &$17.96$ &$-24.16$ &$8.839$ &$-69.25$   &$-$\\
&$1.3199$ &$6.030$ &$5.927$ &$2.929$ &$0.0610$ &$14.97$ &$4.927$ &$17.88$ &$-24.68$ &$8.851$ &$-71.55$   &$-$\\
&$1.2500$ &$5.994$ &$6.021$ &$2.931$ &$0.0623$ &$15.12$ &$4.793$ &$17.83$ &$-25.72$ &$8.938$ &$-76.10$   &$-$\\
&$1.1992$ &$5.993$ &$6.103$ &$2.932$ &$0.0640$ &$15.24$ &$4.657$ &$17.83$ &$-26.85$ &$9.058$ &$-81.19$   &$-$\\
&$1.1160$ &$6.042$ &$6.268$ &$2.937$ &$0.0685$ &$15.50$ &$4.349$ &$17.89$ &$-29.62$ &$9.381$ &$-94.07$   &$-$\\
&$0.9665$ &$6.282$ &$6.712$ &$2.952$ &$0.0835$ &$16.22$ &$3.499$ &$18.21$ &$-38.85$ &$10.44$ &$-141.1$ &$0.1274$\\
&$0.8781$ &$6.493$ &$7.115$ &$2.967$ &$0.0981$ &$16.90$ &$2.791$ &$18.51$ &$-48.44$ &$11.45$ &$-196.2$ &$-0.5157$\\
&$0.8172$ &$6.639$ &$7.482$ &$2.980$ &$0.1111$ &$17.56$ &$2.199$ &$18.70$ &$-57.65$ &$12.30$ &$-254.4$ &$-0.7332$\\
&$0.7604$ &$6.764$ &$7.927$ &$2.995$ &$0.1261$ &$18.49$ &$1.463$ &$18.75$ &$-69.21$ &$13.28$ &$-334.2$ &$-0.8412$\\
&$0.7580$ &$6.769$ &$7.948$ &$2.996$ &$0.1268$ &$18.54$ &$1.423$ &$18.74$ &$-69.78$ &$13.32$ &$-338.4$ &$-0.8437$\\
\hline
\end{tabular}
\end{center}
\caption{Same as Table \ref{EOSA}, but for EOS L.}
\label{EOSL}
\end{table}

\clearpage

\begin{table}
\begin{center}
\begin{tabular}{||p{0.01cm}*{12}{c|}|}
\hline
&$\epsilon_c$ &$\Omega$ &$I$ &$M$ &$T/W$ &$R_e$ &$h_+$ &$h_-$ &$M_2$ &$J$ &$S_3$ &$b$\\
&$10^{15}$~g~cm$^{-3}$ &$10^3$~s$^{-1}$ &$10^{45}$g~cm$^2$ &$M_\odot$ &$-$ &km &km &km &km$^3$ &km$^2$ &km$^4$ &km\\
\hline
\hline
\multicolumn{13}{c}{$M_B=1.551 M_\odot$}\\
\hline
&$0.9950$ &$0.000$ &$-$ &$1.403$ &$0.0000$ &$11.55$ &$0.874$ &$8.737$  &$0.000$ &$0.0000$ &$0.000$     &$-$\\
&$0.9800$ &$2.756$ &$1.368$ &$1.407$ &$0.0124$ &$11.81$ &$-$     &$2.425$  &$-1.965$ &$0.9349$ &$-1.750$   &$-$\\
&$0.9700$ &$3.481$ &$1.396$ &$1.409$ &$0.0202$ &$11.99$ &$-$     &$2.856$  &$-3.203$ &$1.203$ &$-3.673$   &$-0.3370$\\
&$0.9600$ &$4.045$ &$1.425$ &$1.411$ &$0.0280$ &$12.18$ &$-$     &$3.195$  &$-4.464$ &$1.428$ &$-6.064$   &$-0.6044$\\
&$0.9500$ &$4.506$ &$1.455$ &$1.412$ &$0.0356$ &$12.37$ &$-$     &$3.470$  &$-5.723$ &$1.624$ &$-8.852$   &$-0.7155$\\
&$0.9400$ &$4.896$ &$1.486$ &$1.414$ &$0.0432$ &$12.58$ &$-$     &$3.699$  &$-7.002$ &$1.802$ &$-12.03$  &$-0.7689$\\
&$0.9200$ &$5.523$ &$1.551$ &$1.418$ &$0.0581$ &$13.05$ &$-$     &$4.036$  &$-9.608$ &$2.121$ &$-19.50$  &$-0.7935$\\
&$0.9000$ &$6.007$ &$1.621$ &$1.422$ &$0.0727$ &$13.62$ &$-$     &$4.209$  &$-12.31$ &$2.411$ &$-28.49$ &$-0.7644$\\
&$0.8800$ &$6.383$ &$1.696$ &$1.426$ &$0.0870$ &$14.41$ &$-$     &$4.129$  &$-15.14$ &$2.681$ &$-39.17$ &$-0.7089$\\
&$0.8700$ &$6.541$ &$1.737$ &$1.428$ &$0.0943$ &$15.04$ &$-$     &$3.855$  &$-16.64$ &$2.814$ &$-45.32$ &$-0.6744$\\
\hline
\multicolumn{13}{c}{$M_B=2.672 M_\odot$}\\
\hline
&$2.6000$ &$0.000$ &$-$ &$2.205$ &$0.0000$ &$10.12$ &$9.404$ &$9.404$  &$0.000$ &$0.000$ &$0.000$     &$-$\\
&$2.4500$ &$2.820$ &$2.298$ &$2.212$ &$0.0059$ &$10.28$ &$7.707$ &$10.92$ &$-0.905$ &$1.605$ &$-0.526$   &$-$\\
&$2.3000$ &$3.828$ &$2.352$ &$2.217$ &$0.0115$ &$10.45$ &$6.983$ &$11.42$ &$-1.846$ &$2.230$ &$-1.606$   &$-$\\
&$2.1000$ &$5.360$ &$2.451$ &$2.228$ &$0.0244$ &$10.77$ &$5.835$ &$12.24$ &$-4.059$ &$3.254$ &$-5.299$   &$-$\\
&$1.9000$ &$6.791$ &$2.593$ &$2.243$ &$0.0432$ &$11.21$ &$4.628$ &$13.06$ &$-7.604$ &$4.362$ &$-13.71$  &$-$\\
&$1.8000$ &$7.465$ &$2.687$ &$2.253$ &$0.0555$ &$11.50$ &$3.990$ &$13.48$ &$-10.06$ &$4.967$ &$-20.93$ &$-$\\
&$1.7000$ &$8.066$ &$2.799$ &$2.263$ &$0.0693$ &$11.86$ &$3.343$ &$13.86$ &$-13.07$ &$5.592$ &$-31.10$ &$-$\\
&$1.6000$ &$8.605$ &$2.939$ &$2.276$ &$0.0853$ &$12.33$ &$2.660$ &$14.19$ &$-16.82$ &$6.263$ &$-45.57$ &$-$\\
&$1.5000$ &$9.048$ &$3.113$ &$2.290$ &$0.1032$ &$12.97$ &$1.909$ &$14.42$ &$-21.52$ &$6.977$ &$-66.31$ &$-0.2276$\\
&$1.4000$ &$9.384$ &$3.337$ &$2.307$ &$0.1233$ &$14.17$ &$0.767$ &$14.21$ &$-27.54$ &$7.756$ &$-96.55$ &$-0.5089$\\
\hline
\multicolumn{13}{c}{$M_B=2.800 M_\odot$}\\
\hline
&$2.5000$ &$8.310$ &$2.612$ &$2.335$ &$0.0540$ &$10.75$ &$4.968$ &$15.03$  &$-9.015$ &$5.377$ &$-16.70$  &$-$\\
&$2.4000$ &$8.310$ &$2.647$ &$2.336$ &$0.0556$ &$10.86$ &$4.841$ &$15.02$  &$-9.418$ &$5.447$ &$-17.93$  &$-$\\
&$2.3000$ &$8.342$ &$2.686$ &$2.338$ &$0.0579$ &$10.99$ &$4.679$ &$15.01$  &$-9.976$ &$5.549$ &$-19.68$  &$-$\\
&$2.2000$ &$8.457$ &$2.734$ &$2.342$ &$0.0617$ &$11.13$ &$4.450$ &$15.07$  &$-10.82$ &$5.726$ &$-22.32$ &$-$\\
&$2.1000$ &$8.596$ &$2.791$ &$2.346$ &$0.0664$ &$11.30$ &$4.185$ &$15.15$  &$-11.91$ &$5.942$ &$-25.97$ &$-$\\
&$2.0000$ &$8.799$ &$2.860$ &$2.351$ &$0.0728$ &$11.52$ &$3.850$ &$15.27$  &$-13.39$ &$6.233$ &$-31.09$ &$-$\\
&$1.9000$ &$9.023$ &$2.943$ &$2.358$ &$0.0807$ &$11.78$ &$3.467$ &$15.40$  &$-15.26$ &$6.577$ &$-38.03$ &$-$\\
&$1.8000$ &$9.258$ &$3.044$ &$2.367$ &$0.0901$ &$12.10$ &$3.030$ &$15.55$  &$-17.62$ &$6.979$ &$-47.47$ &$-$\\
&$1.7000$ &$9.494$ &$3.168$ &$2.377$ &$0.1013$ &$12.52$ &$2.521$ &$15.68$  &$-20.64$ &$7.449$ &$-60.54$ &$-$\\
&$1.6000$ &$9.705$ &$3.321$ &$2.388$ &$0.1143$ &$13.11$ &$1.894$ &$15.72$  &$-24.45$ &$7.983$ &$-78.60$ &$0.0123$\\
\hline
\end{tabular}
\end{center}
\caption{Same as Table \ref{EOSA}, but for EOS APRb.}
\label{EOSAPRb}
\end{table}

\clearpage

\begin{table}
\begin{center}
\begin{tabular}{||p{0.01cm}*{3}{c|}|}
\hline
&EOS(sequence) & $j_{crit}$ & $\Omega_{crit}/\Omega_{Kepler}$ \\
\hline
\hline
&A(1.44) &$0.39$ &$0.72$ \\
&A(MM) &$0.50$ &$0.89$ \\
\hline
&FPS(1.44) &$0.30$ &$0.57$ \\
&FPS(MM) &$0.50$ &$0.89$ \\
\hline
&L(1.44) &$0.23$ &$0.40$ \\
&L(MM) &$0.52$ &$0.87$ \\
\hline
\end{tabular}
\end{center}
\caption{
Minimum (critical) rotation parameter $j_{crit}$ and corresponding 
ratio of critical angular velocity $\Omega_{crit}$ to Keplerian angular
velocity $\Omega_{Kepler}$, for which the 
matching condition (\ref{Qmatch}) has a real solution.}\label{jcrit}
\end{table}

\begin{table}
\begin{center}
\begin{tabular}{||p{0.01cm}*{3}{c|}|}
\hline
&EOS(sequence) & $\Delta S_{crit}$ & $\Delta S_{Kepler}$ \\
\hline
\hline
&A(1.44) &$-11\%$ &$-4\%$ \\
&A(MM) &$-5\%$ &$-2\%$ \\
&A(SM) &$-2\%$ &$-4\%$ \\
\hline
&AU(1.44) &$-11\%$ &$-5\%$ \\
&AU(MM) &$-9\%$ &$-7\%$ \\
&AU(SM) &$-11\%$ &$-10\%$ \\
\hline
&FPS(1.44) &$-11\%$ &$-4\%$ \\
&FPS(MM) &$-3\%$ &$-2\%$ \\
&FPS(SM) &$-7\%$ &$-4\%$ \\
\hline
&L(1.44) &$-45\%$ &$-4\%$ \\
&L(MM) &$-3\%$ &$-2\%$ \\
&L(SM) &$-12\%$ &$-4\%$ \\
\hline
&APRb(1.4) &$-13\%$ &$-6\%$ \\
&APRb(MM) &$-8\%$ &$-6\%$ \\
&APRb(SM) &$-11\%$ &$-11\%$ \\
\hline
\end{tabular}
\end{center}
\caption{
Relative difference in $S_3$ between the negative
branch of analytic solutions and the numerical solution,
for different EOSs and evolutionary sequences. The difference 
is tabulated for the minimum (critical) rotation rate for
which a real analytic solution exists ($\Delta S_{crit}$) 
and for the model at the mass-shedding limit 
($\Delta S_{Kepler}$).}\label{S3error}
\end{table}

\end{document}